\documentclass[12pt]{article}

\usepackage{natbib}
\usepackage[nottoc,notlof,notlot]{tocbibind} 

\bibpunct{(}{)}{;}{a}{}{;}
\usepackage{breakcites}
\usepackage[breaklinks=true]{hyperref}
\usepackage{graphicx}
\graphicspath{ {images/} }

\usepackage{amsmath}
\usepackage{bbm}
\usepackage{amsfonts}%
\usepackage{amssymb}%

\usepackage{geometry} 
\usepackage{graphicx} 
\usepackage{framed}
\usepackage{caption}
\usepackage{subcaption}

\usepackage{booktabs} 
\usepackage{array} 
\usepackage{paralist} 
\usepackage{verbatim} 

\usepackage{lipsum}
\usepackage{setspace}

\usepackage{lipsum}
\usepackage[singlelinecheck=false]{caption}
\usepackage{tabularx}

\title{\setstretch{1} Testing for Differences in Stochastic Network Structure}

\author{Eric Auerbach\footnote{Department of Economics, Northwestern University. E-mail: eric.auerbach@northwestern.edu. I thank Manuel Arellano, Jon Auerbach, Lori Beaman, Vincent Boucher, Yong Cai, Ivan Canay, Christina Chung, Ivan Fernandez-Val, Eric Gautier, Ben Golub, Bryan Graham, Joel Horowitz, Guido Imbens, Chuck Manski, Angelo Mele, Roger Moon, Aureo de Paula, Sida Peng, Stephen Nei, Mikkel S{\o}lvsten, Alireza Tahbaz-Salehi, Chris Udry, and anonymous referees for helpful suggestions.}} \date{\parbox{\linewidth}{\centering%
  \today\endgraf}} 

\begin{document}
\maketitle
\begin{abstract} \setstretch{1}\noindent
How can one determine whether a treatment, such as the introduction of a social program or trade shock, alters agents' incentives to form links in a network? This paper proposes analogues of a two-sample Kolmogorov-Smirnov test, widely used in the literature to test the null hypothesis of ``no treatment effects,'' for network data. It first specifies a testing problem in which the null hypothesis is that two networks are drawn from the same random graph model. It then describes two randomization tests based on the magnitude of the difference between the networks' adjacency matrices as measured by the $2\to2$ and $\infty\to1$ operator norms. Power properties of the tests are examined analytically, in simulation, and through two real-world applications. A key finding is that the test based on the $\infty\to1$ norm can be substantially more powerful than that based on the $2\to2$ norm for the kinds of sparse and degree-heterogeneous networks common in economics. 
\end{abstract}

\section{Introduction}
This paper proposes analogues of a two-sample Kolmogorov-Smirnov (KS) test for networks. The KS test is a standard way to assess whether two random vectors come from the same distribution and has many applications in economics. One example is detecting distributional treatment effects \citep[see][]{imbensRubin2015}. While tests that compare averages or ranks may ignore key differences between distributions, the KS test compares empirical distribution functions directly, and so with enough data can detect any fixed difference. 

What is an analogous way to detect differences between networks? This paper proposes tests based on operator norms to assess whether two stochastic networks come from the same random graph model. While tests that compare network statistics such as measures of density, clustering, or centrality may ignore key differences between models  \citep[see for instance][]{banerjee2018changes}, the proposed tests, like the KS test for distributions, with enough data can detect any fixed difference. 

Section 2 describes the model, testing problem, and applications. The model is a nonparametric version of a class of dyadic regression models popular in the networks literature. The testing problem is that two networks are drawn from the same model. Applications include tests for network non-stationarity, treatment effects, externalities, and more.


Section 3 outlines a randomization test. The test is based on an implication of the model, that under the null hypothesis the joint distribution of links is invariant to exchanging the weight of a link in one network for its identically indexed counterpart in the other. The test controls size by construction. Power depends on a choice of test statistic.

Section 4 considers two test statistics that produce tests powerful against a large class of alternative hypotheses. The first test statistic is based on the $2\to2$ operator norm (also known as the spectral norm or radius) of the  difference between the networks' adjacency matrices. The second test statistic is based on the $\infty\to1$ operator norm. The tests are easy to implement and quick to compute for networks connecting hundreds of agents. Evaluating their power properties, however, requires new tools from the random matrix theory literature. 

A key result is that while both tests with enough data will detect any fixed difference between models, the test based on the $\infty\to1$ norm can be considerably more powerful when there is nontrivial heterogeneity in the row-variances of the networks' adjacency matrices. Such row-heteroskedasticity may occur when the networks are sparse or have heavy-tailed degree distributions, which characterizes many social and economic networks \citep[see][]{jackson2007meeting}. This is why I recommend that researchers use the test based on the $\infty\to1$ norm when testing for differences between networks in practice. 

Intuitively, the $2\to2$ norm may have low power under row-heteroskedasticity because the weight vector that maximizes this norm places most of its weight on the entries corresponding to the rows of the adjacency matrices with the highest variances. As a result, the test potentially ignores differences between networks that occur in the low-variance rows. The $\infty\to1$ norm addresses the problem by using the $\infty$-vector norm instead of the $2$-vector norm to define the unit weight vector. The entries of the weight vector that maximize this norm take values in $\{-1,1\}$ and so by construction necessarily place the same absolute weight on every entry. Consequently, if there are sufficiently large differences between the low-variance rows, the test based on the $\infty\to1$ norm will detect them. In some sense, this logic behind the $\infty$-vector norm is related to that behind the $1$ or $0$-vector norm penalizations common in the high-dimensional regression literature. Instead of imposing a sparse solution, however, the $\infty$-vector norm imposes a dense one.

Section 5 provides two empirical demonstrations using data from real-world social networks. In both examples, the $\infty\to1$ norm has sufficient power to detect the relevant difference, corroborating theoretical results. Alternatives are less reliable. Additional results, details, and simulation evidence can be found in an online appendix. R code for implementation can be found on my website.\footnote{\url{https://sites.google.com/site/ericjauerbach/}}

\section{Framework}
Sections 2.1 and 2.2 describe the model and testing problem. Section 2.3 provides example applications to testing problems in the network economics literature. 


\subsection{Model}
It is without loss to consider undirected unipartite networks. These networks are defined on a set of $N$ agents referred to as a \emph{community} and indexed by $[N] := \{1,2,...,N\}$. Every pair of agents in a community is endowed with two real-valued random variables, each corresponding to a stochastic social relationship. For example, one weight might correspond to whether two agents are friends, another might give the amount of trade between them, etc. The variable $D_{ij,t}$ for $t = 1,2$ records the realized relationship $t$ between agents $i$ and $j$. The $N\times N$ dimensional symmetric adjacency matrix $D_{t}$ contains $D_{ij,t}$ in the $ij$th and $ji$th entries.

I suppose that the networks to be compared are defined on the same community of agents. In many settings, the community is defined so that this is the case. For instance, to study trade networks economists often define the unit of analysis to be countries and the links to be the amount of trade between countries in a year, even if the agents that actually trade are people and firms that change over time. Testing for differences between networks defined on different communities generally requires more structure concerning exactly how two communities ought to be compared. This is left to future work. 

Directed or bipartite networks are incorporated in the following way. These networks are generally defined on a set of $N_{1}$ agents and $N_{2}$ markets indexed by $[N_{1}]$ and $[N_{2}]$ respectively. Every agent-market pair is endowed with two real-valued random variables, each corresponding to a stochastic social relationship. For example, one weight might correspond to whether the agent is employed in the market, another might give the amount of profit the agent makes in the market, etc. The variable $D^{\star}_{ij,t}$ records the realized relationship $t$ between agent $i$ and market $j$. The $N_{1} \times N_{2}$ dimensional matrix $D^{\star}_{t}$ contains $D^{\star}_{ij,t}$ in the $ijth$ entry. This asymmetric rectangular adjacency matrix is transformed into a symmetric square one by setting 
\begin{align*}
D_{t} = \begin{bmatrix} 0_{N_{1}\times N_{1}} & D^{\star}_{t} \\ \left(D^{\star}_{t}\right)^{T} & 0_{N_{2}\times N_{2}}  \end{bmatrix}
\end{align*}
where $(\cdot)^{T}$ is the transpose operator, $0_{N_{1}\times N_{1}}$ is an $N_{1}\times N_{1}$ matrix of zeros, and $D_{t}$ is a $N\times N$ symmetric matrix with $N = N_{1}+N_{2}$. It follows that the focus on undirected unipartite networks (symmetric and square adjacency matrices) is without loss. 

The entries of $D_{1}$ and $D_{2}$ are assumed be equal to $0$ on the main diagonal (no self-links) and mutually independent above the main diagonal. This independence assumption is common in the dyadic regression literature (see below). The marginal distribution of $D_{ij,t}$ is denoted by $F_{ij,t}$ and the $N\times N$ dimensional matrix $F_{t}$ contains $F_{ij,t}$ in the $ij$th entry.  A generic matrix of distribution functions $F_{t}$ is referred to as a \emph{random graph model}. 


A concrete example of a random graph model is $D_{ij,t} = \mu_{ij,t} + \varepsilon_{ij,t}$, where $\mu_{ij,t} = f_{t}(\alpha_{i,t},\alpha_{j,t},w_{ij,t})$, $\alpha_{i,t}$ is an agent-specific effect, $w_{ij,t}$ are agent-pair attributes, $f_{t}$ is a community link function, and $\varepsilon_{ij,t}$ is an idiosyncratic error that is independently distributed across agent-pairs with marginal distribution $G_{ij,t}$  \cite[see generally][Section 6.3]{graham2019network}. I treat the effects $\{\alpha_{i,t}\}_{i \in [N],t\in[2]}$ and attributes $\{w_{ij,t}\}_{i,j \in [N],t\in[2]}$ as non-stochastic. That is, if these variables are drawn from some distribution, the random graph model is defined conditional on their realization. Such conditioning is standard in the literature.

The only remaining source of randomness are the $\{\varepsilon_{ij,t}\}_{i,j \in [N],t\in[2]}$ and so the $\{D_{ij,t}\}_{i < j \in [N],t\in[2]}$ are independent random variables with marginals given by $F_{ij,t}(s) = G_{ij,t}(s-  \mu_{ij,t}) = G_{ij,t}(s-  f_{t}(\alpha_{i,t},\alpha_{j,t},w_{ij,t}))$. The random graph model $F_{t}$ is thus parametrized by the agent effects, attributes, link function, and distribution of idiosyncratic errors. Informally, if a treatment alters any of these parameters, then the framework characterizes the change as a treatment effect. This is formalized by the statement of the testing problem below. Not every difference in model parameters can be detected using $D_{1}$ and $D_{2}$, however. To constitute evidence against the hypothesis of no treatment effect, the difference in parameters must yield a sufficiently large difference between $F_{1}$ and $F_{2}$. Other notions of a random graph model may imply other definitions of a treatment effect. Their study is left to future work. 





\subsection{Testing problem}
The problem considered in this paper is to test the null hypothesis 
\begin{align*}
H_{0}: F_{ij,1} = F_{ij,2} \text{ for every } i,j \in [N]
\end{align*}
against the alternative 
\begin{align*}
H_{1}: F_{ij,1} \neq F_{ij,2} \text{ for some } i,j \in [N].
\end{align*}
$H_{0}$ is the hypothesis that $D_{1}$ and $D_{2}$ are drawn from the same random graph model.



For the concrete example of Section 2.1, the problem is equivalent to testing the hypothesis that $G_{ij,1}(s-  f_{1}(\alpha_{i,1},\alpha_{j,1},w_{ij,1})) = G_{ij,2}(s-  f_{2}(\alpha_{i,2},\alpha_{j,2},w_{ij,2}))$ for every  $s \in \mathbb{R}$ and $i,j \in [N]$. The hypothesis may be false whenever the two random graph models have different agent-specific effects, agent-pair attributes, community link functions, or distributions of idiosyncratic errors. Distinguishing between these parameters generally requires more structure. For example, if the  $\{\varepsilon_{ij,t}\}_{i \in [N_{T}],t\in[2]}$ are identically distributed, $\alpha_{i,1} = \alpha_{i,2}$, and $w_{ij,1} = w_{ij,2}$ for every $i,j \in [N]$, then the problem reduces to a test of whether the link functions $f_{1}$ and $f_{2}$ are the same. If the $\{\varepsilon_{ij,t}\}_{i \in [N_{T}],t\in[2]}$ are identically distributed, $f_{1} = f_{2}$, and $w_{ij,1} = w_{ij,2}$ for every $i,j \in [N]$, then the problem reduces to a test of whether $\alpha_{i,1} = \alpha_{i,2}$ for every $i \in [N]$. See Online Appendix Section B.2 for a discussion.

The logic behind this test is that any change in the distribution of idiosyncratic errors, agent fixed effects, link function, etc. reflects different incentives for agents to form or report links. Differences due to the idiosyncratic errors reflect only policy-irrelevant statistical noise. As previously motivated, the test is specifically designed to detect a large class of differences between two random graph models. Other testing problems tailored to detect more specific differences are discussed in the applications and extensions below. 


Related testing problems have also been considered in the literature. \cite{ghoshdastidar2017two1,ghoshdastidar2017two2} propose a test statistic based on the $2\to2$ norm for a different testing problem. A literature on random dot product graph models \citep[see][]{tang2017nonparametric,nielsen2018multiple} relies on a low-dimensional dot-product structure. A related application of randomization-based inference to networks is tests for interference \cite[see for instance][]{aronow2012general,athey2018exact}. Rather than study the influence of a treatment on network structure, this literature studies the influence of a network on agents' exposure to treatment.

\subsection{Example applications and extensions}
I sketch six applications and three extensions of the framework to testing problems in the network economics literature. Details can be found in Online Appendix Section D. 

\subsubsection{Application 1: a test of link stationarity}
 \cite{goyal2006economics} observe co-authorships between economists over time and argue that the profession has become more interconnected in response to new research technologies such as the internet. The above framework can be used to test whether the differences over time are statistically significant. The first example in Section 5 demonstrates this application to testing link stationarity. 
 
 \subsubsection{Application 2: a test for link heterogeneity}
\cite{banerjee2013diffusion} collect data on a dozen social and economic ties between villagers. \cite{jackson2017economic} suggest that this data may ``encode richer information than simply identifying whether two people are close or not.'' The above framework can be used to test whether the differences between the networks induced by different survey questions are statistically significant. The second example in Section 5 demonstrates this application to testing link homogeneity. 

\subsubsection{Application 3: a test of no treatment effects}
\cite{rose2004we} finds that country participation in trade agreements such as the WTO does not significantly alter international trade. The above framework can be used to test whether program participation leads to a statistically significant change in network structure.

\subsubsection{Application 4: a test for endogenous link formation}
 \cite{goldsmith2013social} consider a model of student GPA and link formation, and test whether a determinant of GPA also drives link formation in the network. They propose a one-sample parametric test for such endogenous link formation. The above framework can be used for an alternative two-sample nonparametric test.  

\subsubsection{Application 5: a test of link reciprocity}
\cite{calvo2009peer} specify a model of network peer effects in which any nomination of a friendship from one student to another indicates a social tie between both students. The above framework can be used to test for reciprocity or symmetry in link nominations.\footnote{I thank Vincent Boucher for suggesting the example.}

\subsubsection{Application 6: a test for network externalities}
\cite{pelican2020optimal} consider a model of link formation in which the decision for two agents to link may depend on other links that have been realized in the network. They propose a one-sample parametric test for such network externalities. The above framework can be used for an alternative two-sample nonparametric test.


\subsubsection{Extension 1: a completely randomized experiment}
\cite{banerjee2018changes} collect data on connections between villagers in a number of villages before and after a microfinance agency offers loans to villagers in some but not all of the villages. They find that the program disincentivizes the formation of certain types of connections. The above framework can be modified to test whether the observed differences between the treatment and control villages are statistically significant.  

\subsubsection{Extension 2: a one-sample test of independence}
\cite{fafchamps2007risk} study a risk-sharing network and argue that the surveyed links are not related to the respondents' occupations. The above framework can be modified to test whether the network connections and respondent occupations are statistically independent.

\subsubsection{Extension 3: a one-sample specification test}
\cite{jackson2007meeting} argue that real-world social networks have features that are not explained by an Erd\"os-Renyi model of link formation. The above framework can be modified to test whether network data can be explained by a specific random graph model.

\section{Randomization procedure}
I outline a randomization procedure to construct tests for the problem of Section 2 \citep[see][Chapter 15]{lehmann2006testing}. The procedure takes as given a test statistic $T(D_{1},D_{2})$. Any real-valued function of $D_{1}$ and $D_{2}$ can be used to construct the test statistic. Example test statistics include differences in centrality measures such as agent degree, eigenvector centrality, or clustering. Certain centrality measures may direct the power of the test towards specific alternatives.  



 
For any positive integer $R$, let $\{\rho_{ij}^{r}\}_{i > j \in [N], r \in [R]}$ be a collection of ${N \choose 2} \times R$ independent Bernoulli random variables with mean $1/2$. Define $\rho_{ij}^{r} = \rho_{ji}^{r}$ if $i < j$. Then for each $r \in [R]$, the randomized $N \times N$ adjacency matrices $D_{1}^{r}$ and $D_{2}^{r}$ are generated by exchanging $D_{ij,1}$ and $D_{ij,2}$ whenever $\rho_{ij}^{r}$ equals $1$. That is, 
\begin{align*}
D_{ij,1}^{r} &= D_{ij,1}\rho_{ij}^{r} + D_{ij,2}(1-\rho_{ij}^{r})\\
D_{ij,2}^{r} &= D_{ij,1}(1-\rho_{ij}^{r}) + D_{ij,2}\rho_{ij}^{r}
\end{align*}
where $D_{ij,1}^{r}$ is the $ij$th entry of $D_{1}^{r}$. For any $\alpha \in [0,1]$, the proposed $\alpha$-sized test based on $T(D_{1},D_{2})$ rejects $H_{0}$ if
\begin{align*}
(R+1)^{-1}\left(1 + \sum_{r\in[R]}\mathbbm{1}\left\{T(D_{1}^{r},D_{2}^{r}) \geq T(D_{1},D_{2})\right\} \right) \leq \alpha
\end{align*}
and fails to reject $H_{0}$ otherwise.

Since  $(D_{1},D_{2})$ and $(D_{1}^{r},D_{2}^{r})$ have the same distribution under $H_{0}$,  \cite{lehmann2006testing}, Theorem 15.2.1 implies the following. When $H_{0}$ is true the probability of an (incorrect) rejection does not exceed $\alpha$, or
\begin{align*}
P\left((R+1)^{-1}\left(1 + \sum_{r\in[R]}\mathbbm{1}\left\{T(D_{1}^{r},D_{2}^{r}) \geq T(D_{1},D_{2})\right\} \right) \leq \alpha | H_{0} \text{ is true }\right) \leq \alpha.
\end{align*}
This is true for any test statistic $T(D_{1},D_{2})$.  When the researcher has a collection of network statistics, the tests can be combined in the usual way. One can also test whether any number of adjacency matrices are drawn from the same random graph model by permuting all of the corresponding entries.

\section{Two tests based on operator norms}
Two tests based on operator norms are proposed in Section 4.1. Large sample power properties of the tests are characterized in Section 4.2 and Online Appendix Section B. 

\subsection{Specification}
The randomization procedure of Section 3 produces a test for the problem of Section 2 that controls the probability of (incorrectly) rejecting the null hypothesis when it is true using any test statistic. However, not every statistic produces a test that is powerful in that it tends to (correctly) reject the null hypothesis when it is false. This section proposes two statistics that have power against a large class of alternative hypotheses. The statistics are based on $p\to q$ operator norms. 

For any $p,q \geq 1$, the test statistic based on the $p\to q$ operator norm is given by
\begin{align*}\label{pq}
T_{p\to q}(D_{1},D_{2}) = \max_{s \in \mathbb{R}}\max_{\varphi : ||\varphi||_{p} = 1}||\left[\mathbbm{1}\{D_{1} \leq s\}-\mathbbm{1}\{D_{2} \leq s\}\right]\varphi||_{q}
\end{align*} 
where $||\cdot||_{p}$ refers to the vector $p$-norm, $\varphi$ is a $N$-dimensional column vector with real-valued entries, and for any real number $s$ and matrix $X$, $\mathbbm{1}\{X \leq s\}$ contains $1$ in the $ij$th entry if $X_{ij} \leq s$ and $0$ otherwise. The test of $H_{0}$ based on $T_{p\to q}$ is as described in Section 3.  

When the entries of $D_{1}$ and $D_{2}$ are $\{0,1\}$-valued,
\begin{align*}
T_{p\to q}(D_{1},D_{2}) = \max_{\varphi : ||\varphi||_{p} = 1}||\left(D_{1}-D_{2} \right)\varphi||_{q}
\end{align*}
which is the $p\to q$ operator norm of the entry-wise difference between two adjacency matrices. Intuitively, $T_{p\to q}\left(D_{1},D_{2}\right)$ compares the collection of weighted degree distributions of $D_{1}$ and $D_{2}$,  indexed by the weight vector $\varphi$ and given by $\{\sum_{j\in[N]}D_{ij,t}\varphi_{j}\}_{i \in [N], \varphi: ||\varphi||_{p} =1}$. This weighted degree distribution function is a matrix analogue to the empirical distribution function for vectors. Instead of measuring the number of entries that fall below a point on the real line, it measures the magnitude of connections from each agent to the community as weighted by $\varphi$. 

Not every $p\to q$ operator norm is either computable or produces a test that has power against a nontrivial class of alternatives. This is why I focus on two choices of $p$ and $q$.

The first test statistic is based on the $2\to2$ operator norm (also known as the spectral norm or spectral radius)
\begin{align*}
T_{2\to2}(D_{1},D_{2}) = \max_{s \in \mathbb{R}}\max_{\varphi: \sum_{t}\varphi_{t}^{2} = 1}\sqrt{\sum_{i \in [N]}\left(\sum_{j \in [N]}\left[\mathbbm{1}\{D_{ij,1} \leq s\}-\mathbbm{1}\{D_{ij,2} \leq s\}\right]\varphi_{j}\right)^{2}}.
\end{align*}
The $2\to2$ norm is a natural first choice because it is straightforward to compute in $O(N^3)$ time and its statistical properties have been well studied in the random matrix theory literature. However, as I demonstrate below, the resulting test may have low power under row-heteroskedasticity: nontrivial variation in the row-variances of the adjacency matrices. Intuitively, the problem is that the weight vector $\varphi$ that maximizes the program may place excessive weight on the high-variance rows of $\left[\mathbbm{1}\{D_{1} \leq s\}-\mathbbm{1}\{D_{2} \leq s\}\right]$ (see below). To address this problem, I consider a second test statistic. 

The second test statistic is based on the $\infty\to1$ operator norm
\begin{align*}
T_{\infty\to1}(D_{1},D_{2}) = \max_{s \in \mathbb{R}}\max_{\varphi: \max_{t \in [N]}\left|\varphi_{t}\right| = 1}\sum_{i \in [N]}\left|\sum_{j \in [N]}\left[\mathbbm{1}\{D_{ij,1} \leq s\}-\mathbbm{1}\{D_{ij,2} \leq s\}\right]\varphi_{j}\right|.
\end{align*}

The logic behind this test statistic is that the weight vector $\varphi$ that maximizes this problem necessarily places the same absolute weight on every entry and so is less sensitive to row-heteroskedasticity. Unfortunately, computing this norm is NP-hard. The proposed test is instead based on the semidefinite approximation
\begin{align*}
S_{\infty\to1}(D_{1},D_{2}) &= \frac{1}{2}\max_{s \in \mathbb{R}}\max_{X \in \mathcal{X}_{2N}}\left<\begin{bmatrix} 0_{N\times N} & \Delta(s) \\ \Delta(s) & 0_{N\times N}  \end{bmatrix},X\right>
\end{align*}
where $\Delta(s) = \mathbbm{1}\{D_{1} \leq s\}-\mathbbm{1}\{D_{2} \leq s\}$, $\left<\cdot\right>$ is the inner product operator (i.e. $\left<X,Y\right> = \sum_{i=1}^{2N}\sum_{j=1}^{2N}X_{ij}Y_{ij}$), and $\mathcal{X}_{2N}$ is the set of all $2N \times 2N$ positive semidefinite matrices with diagonal entries equal to 1, see generally \cite{alon2006approximating}. This statistic can be computed in $O\left(N^{3.5}\right)$ time using programs available in many statistical software packages. Its substitution is justified by the fact that
\begin{align*}
T_{\infty\to1} \leq S_{\infty\to1} \leq K T_{\infty\to1}
\end{align*}
for any $D_{1}$, $D_{2}$, and $N$  with $K = \frac{\pi}{2\ln{\left(1 + \sqrt{2}\right)}} \leq 1.783$. A derivation of the semidefinite approximation and justification of the inequalities can be found in Appendix Section A.2. 


\subsection{Consistency}
As discussed in Section 3, the $\alpha$-sized tests based on $T_{2\to 2}$ and $S_{\infty\to 1}$ (incorrectly) reject the null hypothesis when it is true with probability less than $\alpha$. This section provides conditions such that the tests (correctly) reject the null hypothesis when it is false. Specifically, it defines a class of sequences of alternative hypotheses such that the power of the tests tend to one uniformly over the class. Each sequence describes a collection of models and tests indexed by $N \in \mathbb{N}$. Each model is as described in Section 2. Each test is as described in Section 3. The parameters $F_{1}$, $F_{2}$, $R$, and $\alpha$ may all vary with $N$ subject to restrictions below. Limits are with $N \to \infty$. The statement of the results mirror those for the KS test in Chapter 14.2 of \cite{lehmann2006testing}, although to my knowledge the underlying arguments are not related in any meaningful way. 


\subsubsection{Assumptions and constructions}

The following assumptions on the size of the test are imposed. 
\begin{flushleft}
\textbf{Assumptions:} 
\begin{align*}
-\ln(\alpha) = O\left(\ln(N)\right) \text{ and } \alpha R \geq 2 \hspace{5mm} \square
\end{align*}
\end{flushleft}
\vspace{2mm}
I do not believe either to be restrictive in practice. The first is that the size of the test is not exponentially small relative to the number of agents. The second is that the size of the test is larger than twice the inverse of the number of simulations. They are required because if $\alpha$ is too small, the test will mechanically fail to reject $H_{0}$ regardless of the choice of $T$ or difference between $F_{1}$ and $F_{2}$. They follow if $\alpha$ is fixed and $R$ tends to infinity with $N$. 


The following constructions are used.
\begin{flushleft}
\textbf{Constructions:} 
\begin{align*}
\Delta(s) &= \mathbbm{1}\{D_{1} \leq s\}-\mathbbm{1}\{D_{2} \leq s\}\\
\nu_{ij}(s) &= F_{ij,1}(s) + F_{ij,2}(s) - 2F_{ij,1}(s)F_{ij,2}(s)  \\
\tau &= \max_{s \in \mathbb{R}}\max_{i \in [N]}\sqrt{\sum_{j \in [N]}\nu_{ij}(s)} \\
\sigma &= \max_{s \in \mathbb{R}}\sum_{i \in [N]}\sqrt{\sum_{j \in [N]}\nu_{ij}(s)} \\
T_{2\to2}(F_{1},F_{2}) &= \max_{s \in \mathbb{R}}\max_{\varphi: ||\varphi||_{2} = 1}||\left(F_{1}(s)-F_{2}(s)\right)\varphi||_{2} \\
T_{\infty\to1}(F_{1},F_{2}) &= \max_{s \in \mathbb{R}}\max_{\varphi: ||\varphi||_{\infty} = 1}||\left(F_{1}(s)-F_{2}(s)\right)\varphi||_{1}  \hspace{5mm} \square
\end{align*}
\end{flushleft}
\vspace{2mm}

In words, $\nu_{ij}(s)$ is the variance of $\Delta_{ij}(s)$, $\sqrt{\sum_{j \in [N]}\nu_{ij}(s)} $ is the root of the $i$th row-variance of $\Delta(s)$, $\max_{i \in [N]}\sqrt{\sum_{j \in [N]}\nu_{ij}(s)}$ and $\sum_{i \in [N]}\sqrt{\sum_{j \in [N]}\nu_{ij}(s)}$ are the maximum and average root-row-variance of $\Delta(s)$ respectively, and $\tau$ and $\sigma$ are the maximum of the maximum and average root-row-variances taken over $s$. Under $H_{0}$ and certain regularity conditions, $T_{2\to2}(D_{1},D_{2})$ is proportional to $\tau$ and $T_{\infty\to1}(D_{1},D_{2})$ is proportional to $\sigma$ with high probability. This is demonstrated by Lemmas 1 and 2 in Appendix Section A. 

$T_{2\to2}(F_{1},F_{2})$ and $T_{\infty\to1}(F_{1},F_{2})$ are the test statistics applied to the (matrix of) distribution functions $F_{1}$ and $F_{2}$. These metrics quantify the extent to which $H_{0}$ is violated. Larger values correspond to more extreme violations. Under certain regularity conditions, $T_{2\to2}(F_{1},F_{2})$ large relative to $\tau$ or $T_{\infty\to1}(F_{1},F_{2})$ large relative to $\sigma$ eventually results in a rejection of $H_{0}$. This is the content of Theorems 1 and 2 below. 


\subsubsection{Consistency}
The main consistency result for the test based on the $2\to2$ norm is given by Theorem 1.
\vspace{2mm}
\begin{flushleft}
\textbf{Theorem 1}: The power of the $\alpha$-sized test that rejects $H_{0}$ whenever 
\begin{align*}
(R+1)^{-1}\left(1 + \sum_{r\in[R]}\mathbbm{1}\{T_{2\to2}(D_{1}^{r},D_{2}^{r}) \geq T_{2\to2}(D_{1},D_{2})\} \right) \leq \alpha
\end{align*}
tends to one uniformly over all alternatives that satisfy $T_{2\to2}(F_{1},F_{2})/\tau \to \infty$ and  $\tau/\sqrt{\ln(N)} \to \infty$. That is, 
\begin{align*}
\inf_{\substack{F_{1},F_{2}: T_{2\to2}(F_{1},F_{2})/\tau \to \infty \\ \text{ and } \tau/\sqrt{\ln(N)} \to \infty}}P\left((R+1)^{-1}\left(1 + \sum_{r\in[R]}\mathbbm{1}\{T_{2\to2}(D_{1}^{r},D_{2}^{r}) \geq T_{2\to2}(D_{1},D_{2})\} \right) \leq \alpha\right)\to 1.  \hspace{5mm} \square
\end{align*}
\end{flushleft} 
\vspace{2mm}

The hypothesis of Theorem 1 has two rate conditions. The first rate condition is that $T_{2\to2}(F_{1},F_{2})/\tau \to \infty$. This condition implies that the size of the violation of $H_{0}$ (as given by $T_{2\to2}(F_{1},F_{2})$) exceeds the magnitude of the test statistic $T_{2\to2}(D_{1},D_{2})$ under $H_{0}$ (which is on the order of $\tau$) with high probability. When $F_{t}$ is sufficiently dense in the sense that for some $s \in \mathbb{R}$, $F_{ij,t}(s)$ is uniformly bounded away from $0$ and $1$, it follows from $\sqrt{N}T_{2\to2}(F_{1},F_{2}) \to \infty$. 

The second rate condition is that $\tau/\sqrt{\ln(N)} \to \infty$. This condition is sufficient for the reference distribution generated by $\{T_{2\to2}(D_{1}^{r},D_{2}^{r})\}_{r\in [R]}$ to concentrate below $\tau$. It is satisfied if $F_{t}$ is sufficiently dense in the sense that, for some $s \in \mathbb{R}$, $\frac{N}{\ln(N)}F_{ij,t}(s)$ and $\frac{N}{\ln(N)}(1-F_{ij,t}(s))$  are uniformly bounded away from $0$. In other words, agents have on expectation at least $\ln(N)$ connections. This suggests that the test based on the $2\to2$ norm may be poorly suited for networks in which agents have on expectation only a bounded number of connections. 

The main result for the test based on the $\infty\to1$ norm is given by Theorem 2.
\vspace{2mm}
\begin{flushleft}
\textbf{Theorem 2}: The power of the $\alpha$-sized test that rejects $H_{0}$ whenever 
\begin{align*}
(R+1)^{-1}\left(1 + \sum_{r\in[R]}\mathbbm{1}\{S_{\infty\to1}(D_{1}^{r},D_{2}^{r}) \geq S_{\infty\to1}(D_{1},D_{2})\} \right) \leq \alpha
\end{align*}
tends to one uniformly over all alternatives that satisfy $T_{\infty\to1}(F_{1},F_{2})/\sigma \to \infty$ and $\sigma/\sqrt{\ln(N)} \to \infty$. That is, 
\begin{align*}
\inf_{\substack{F_{1},F_{2}: T_{\infty\to1}(F_{1},F_{2})/\sigma \to \infty \\ \text{ and } \sigma/\sqrt{\ln(N)} \to \infty}}P\left((R+1)^{-1}\left(1 + \sum_{r\in[R]}\mathbbm{1}\{S_{\infty\to1}(D_{1}^{r},D_{2}^{r}) \geq S_{\infty\to1}(D_{1},D_{2})\} \right) \leq \alpha\right)\to 1. \hspace{5mm} \square
\end{align*}
\end{flushleft} 
\vspace{2mm}

The two rate conditions in the hypothesis of Theorem 2 are similar to those in the hypothesis of Theorem 1. The first is that $T_{\infty\to1}(F_{1},F_{2})/\sigma \to \infty$ (or equivalently, that $S_{\infty\to1}(F_{1},F_{2})/\sigma \to \infty)$. This condition implies that the size of the violation of $H_{0}$ (as given by $T_{\infty\to1}(F_{1},F_{2})$) exceeds the magnitude of the test statistic under $H_{0}$ (which is on the order of $\sigma$) with high probability. When $F_{t}$ is dense in the sense that for some $s \in \mathbb{R}$, $F_{ij,1}(s)$ is uniformly bounded away from $0$ and $1$, it also follows from $\sqrt{N}T_{\infty\to1}(F_{1},F_{2}) \to \infty$. 


The second rate condition is that $\sigma/\sqrt{\ln(N)} \to \infty$. This condition is sufficient for the reference distribution generated by $\{S_{\infty\to1}(D_{1}^{r},D_{2}^{r})\}_{r \in [R]}$ to concentrate below $\sigma$. It is satisfied if for some $s \in \mathbb{R}$, $\frac{N^3}{\ln(N)}F_{ij,t}(s)$ and $\frac{N^3}{\ln(N)}(1-F_{ij,t}(s))$ are uniformly bounded away from $0$. In contrast to the hypothesis of Theorem 1, it is sufficient that agents have on expectation at least $\ln(N)/N^2$ connections, which covers settings in which agents have a bounded number of connections. The condition is unlikely to be restrictive in practice. 


The results predict two scenarios in which the test based on the $\infty\to1$ norm is potentially more powerful than that based on the $2\to2$ norm, in the sense that the hypothesis of Theorem 2 is satisfied but that of Theorem 1 is not. The first scenario is network sparsity. An example of this is when $F_{1}\wedge (1-F_{1})$ and $F_{2}\wedge(1-F_{2})$ are uniformly on the order of $1/N$. One can verify that $\tau/\sqrt{\ln(N)} \to 0$ but $ \sigma/\sqrt{\ln(N)} \to \infty$. The second scenario is degree-heterogeneity. An example of this is when, for some small positive integer $K$, $F_{ij,1} = F_{ij,2}$ is on the order of a constant if $i\wedge j \leq K$ but $F_{ij,1} \neq F_{ij,2}$ is on the order of $1/\sqrt{N}$ when $i\wedge j >  K$. One can verify that  $T_{2\to2}(F_{1},F_{2})/\tau \to 0$ but $T_{\infty\to1}(F_{1},F_{2})/\sigma \to \infty$.\footnote{Alternatively, the test based on the $2\to2$ norm is potentially more powerful under degree-heterogeneity when the differences between $F_{1}$ and $F_{2}$ occur in a small number of high-variance rows, although making inferences based on a small number of high-variance observations is not generally recommended.}  Corroborating simulation evidence can be found in Online Appendix Section C.

 Additional results about the near-optimality of the rate conditions, pointwise consistency, and nonasymptotic bounds on power can be found in Online Appendix Section B. 
 
\section{Two empirical demonstrations}
I provide two empirical demonstrations using publicly available data. In both settings the randomization test based on the $\infty\to1$ norm is sufficiently powerful to detect the relevant difference in network structure. Alternatives are less reliable. 

The first is a test of network stationarity as described in Application 1 of Section 2.3. A sample of high school students are surveyed annually about their social connections. The networks appear to be less connected and more clustered over time, potentially because the students place increasing value on having friends in common as they age. The problem is to test whether the observed changes in reported relationships are statistically significant.

The data comes from the ``Teenage Friends and Lifestyle Study'' \citep*[see][]{michell1996peer} in which the researchers survey 160 Scottish students about friendship links during their second through fourth years of secondary school.\footnote{ The data can be found at \url{https://www.stats.ox.ac.uk/~snijders/siena/Glasgow_data.htm}.} This example uses the social network surveyed from the first and third waves when the students are respectively 13 and 15 years old. Only students who appear in all waves are included, yielding a sample size of $N = 129$. 

Descriptive network statistics are provided in Panel A of Table 1. The table describes the means and standard deviations of four network statistics: the sequence of agent degrees $\{\sum_{j \in [N]}D_{ij}\}_{i \in [N]}$, eigenvector centralities, clustering coefficients $\frac{\sum_{i,j,k \in [N]}D_{ij}D_{ik}D_{jk}}{\sum_{i,j,k \in [N]}D_{ij}D_{ik}}$, and diameters of the connected component. The table indicates that while the total number of links appear to be roughly the same for both networks, the second network is less connected and more clustered than the first. Table 2 demonstrates that these differences are unlikely to be generated by the same random graph model. 

Panel A of Table 2 reports the p-value of the randomization test proposed in Section 3 using seven test statistics. The first five test statistics are the absolute difference in average degree, the mean squared difference in agent degrees, the mean squared difference in eigenvector centralities, absolute difference in clustering coefficients, and absolute difference in diameters of the two networks. The last two test statistics are $T_{2\to2}$ and $S_{\infty\to1}$. Panel A indicates that the differences in the clustering and diameter are unlikely to be generated by the same random graph model. The implausibility of the null hypothesis is also clearly indicated by the tests based on $T_{2\to2}$ and  $S_{\infty\to1}$. 

An alternative way to detect differences in network structure is a regression-based approach where one computes a vector of network statistics for both networks, specifies a regression model in which the network statistics depend on a constant, an indicator for one of the networks, and an idiosyncratic error, and conducts a Wald test \citep[see for example][Section 3]{banerjee2018changes}.\footnote{I thank an anonymous referee for suggesting the comparison. Formally, if $S_{it}$ is a network statistic associated with agent $i$ in network $t$ such as $i$'s degree or eigenvector centrality, then the test is based on the linear model $S_{it} = \alpha + \beta \mathbbm{1}_{t = 2} + \varepsilon_{it}$. It is assumed that the errors $\{\varepsilon_{it}\}_{i \in [N], t \in [2]}$ are independent and identically distributed with normal marginals. The null hypothesis is $\beta = 0$.} Panel A of Table 3 reports the OLS point estimates and p-values for three network statistics on the left-hand side: agent degree, eigenvector centrality, and agent clustering  $\left\{\frac{\sum_{j,k \in [N]}D_{ij}D_{ik}D_{jk}}{\sum_{j,k \in [N]}D_{ij}D_{ik}}\right\}_{i \in [N]}$. In all three regressions, the coefficient in front of the network indicator is not statistically significant: the regression-based tests do not detect the change in structure of the Glasgow networks.  

The second demonstration is a test for link heterogeneity as described in Application 2 in Section 2.3. Households in a village are surveyed about multiple types of relationships. Different survey questions appear to reveal information about different types of connections between agents. The problem is to test whether the observed differences in the network structure induced by the different survey questions are statistically significant. 

The data comes from \cite{banerjee2013diffusion}, who survey information about a dozen social and economic connections between households for $75$ villages in rural India.\footnote{ The data can be found at \url{https://hdl.handle.net/1902.1/21538}.} This demonstration uses data from the $77$ households in village $10$ and compares the social network in which two households are linked if a member of one of the households indicates that they ``engage socially'' with a member of the other household to the economic network in which two households are linked if a member of one of the households indicates that they ``borrow money from,'' ``borrow kerosene or rice from,'' ``lend kerosene or rice to,'' or lend money to'' a member of the other household. 

Panel B of Table 1 describes the same statistics as Panel A, but for this second application. It shows two main differences between the social and economic networks. The first difference is that the surveyed households have on average approximately one more economic link than social link. The second difference is that there is more clustering in the economic network. Table 2 demonstrates that these differences are unlikely to be generated by the same random graph model. 

Panel B of Table 2 describes the same tests as Panel A, but for this second application. It indicates that the differences between the average degrees and clustering coefficients of the two networks are unlikely to be explained by the null hypothesis. However, this difference would not be detected by a reasonably-sized test based on the $2\to2$ norm because $T_{2\to2}(D_{1},D_{2})$ is in the third quartile of its reference distribution. On the other hand,  $S_{\infty\to1}$ is firmly in the upper decile of its reference distribution, and so this test statistic provides evidence against the null hypothesis. 

Panel B of Table 3 describes the same regressions as Panel A, but for this second application. The results from the tests based on agent degree and clustering also indicate a difference in the structure of the social and economic networks. The strength of this evidence, however, depends crucially on the regression model assumptions of linearity, normality, homoskedasticity, etc.

\section{Conclusion}
This paper considers a two-sample testing problem where the null hypothesis is that two networks are drawn from the same random graph model. Two tests based on the magnitude of the difference between the networks' adjacency matrices as measured by the $2\to2$ and $\infty\to1$ operator norms are proposed. Both tests with enough power can detect any fixed difference between random graph models, however, the test based on the $\infty\to1$ is shown to be substantially more powerful for the kinds of sparse and degree-heterogeneous networks common in economics.

\begin{table}[]
\centering
\title{Table 1: Descriptive Statistics} \\ \vspace{5mm}
\begin{tabular}{ccccccccc}
 \toprule \toprule
Panel A: &&Agent  & Eigenvector  & Clustering & Diameter \\
Glasgow Networks &&Degree  & Centrality  & Coefficient &    \\ \midrule
First\\ Wave \\
&Mean    &3.49     &0.09  &0.35  &14.00   \\
&SD        &1.69     &0.20    &0.00  &0.00   \\
Third\\ Wave&&&&&&&&\\
&Mean    &3.64     &0.07   &0.42 &20.00  \\
&SD        &1.80     &0.21     &0.00 &0.00   \\
 \bottomrule
   \toprule
Panel B: &&Agent  & Eigenvector  & Clustering & Diameter \\
India Networks &&Degree  & Centrality  & Coefficient &    \\ \midrule
Social\\ Network \\
&Mean    &3.92     &0.26  &0.13  &6.00   \\
&SD       &3.17    &0.22    &0.00  &0.00   \\
Economic\\ Network&&&&&&&&\\
&Mean    &4.94     &0.27   &0.19 &5.00  \\
&SD        &3.65     &0.21   &0.00 &0.00   \\
 \bottomrule
\end{tabular}
\normalsize
\newline
\begin{flushleft} \footnotesize \linespread{.75} Panel A compares two networks surveyed from 129 students in the ``Teenage Friends and Lifestyle Study.'' Panel B compares two networks from village 10 in \cite{banerjee2013diffusion}. Both panels describe the means and standard deviations for four measures of network structure: the sequence of agent degrees, eigenvector centralities, clustering coefficients, and diameters of the largest connected component. \normalsize \end{flushleft}
\end{table}

\begin{table}[]
\centering
\title{Table 2: Randomization Tests} \\ \vspace{5mm}
\begin{tabular}{ccccccccc}
 \toprule \toprule
Panel A: &&Average &  Agent  & Eigenvector  & Clustering & Diameter & $2\to2$    & $\infty\to1$ \\ 
Glasgow Networks & &Degree  & Degree   & Centrality  & Coefficient  &  & Norm &Norm \\ \midrule
p-value &&&&&&&&\\
&&0.50   &0.89 &0.19 &0.01  &0.06    &0.01 &0.01\\
 \bottomrule
  \toprule
Panel B: &&Average &  Agent  & Eigenvector  & Clustering & Diameter & $2\to2$    & $\infty\to1$ \\ 
India Networks & &Degree  & Degree   & Centrality  & Coefficient  &  & Norm &Norm \\ \midrule
p-value &&&&&&&&\\
&&0.00   &0.58 &0.73 &0.05   &0.55 &0.25 &0.05\\
 \bottomrule
\end{tabular}
\normalsize
\newline
\begin{flushleft} \footnotesize \linespread{.75} Panel A compares two networks surveyed from 129 students in the ``Teenage Friends and Lifestyle Study.'' Panel B compares two networks from village 10 in \cite{banerjee2013diffusion}. Both panels describe p-values for randomization tests based on the absolute difference in average degree, the mean squared difference in agent degrees, the mean squared difference in eigenvector centralities, absolute difference in clustering coefficients, absolute difference in diameters, $2\to2$ norm of the entry-wise differences between the two networks' adjacency matrices, and semidefinite approximation to the $\infty\to1$ norm of the entry-wise differences between the two networks' adjacency matrices. For all tests, $R = 10,000$. \normalsize \end{flushleft}
\end{table}

\begin{table}[]
\centering
\title{Table 3: Regressions} \\ \vspace{5mm}
\begin{tabular}{cccccc}
 \toprule \toprule
Panel A: & &  Agent  & Eigenvector  & Agent \\ 
Glasgow Networks &  & Degree   & Centrality  & Clustering   \\ \midrule
intercept &&&&&\\
&   &3.48 &0.09 &0.21 \\
&   &(0.00) &(0.00) &(0.00) \\
coefficient &&&&&\\
&   &0.12  &-0.02 &-0.03 \\
&   &(0.58)  &(0.34)  &(0.15) \\
 \bottomrule
  \toprule
Panel B: &&   Agent  & Eigenvector  & Agent \\ 
India Networks &   & Degree   & Centrality  & Clustering  \\ \midrule
intercept &&&&&\\
&&3.92 &0.26 &0.03  \\
&&(0.00) &(0.00) &(0.00) \\
coefficient &&&&&\\
&&1.01 &0.01 &0.02 \\
&&(0.07) &(0.76) &(0.01) \\
 \bottomrule
\end{tabular}
\normalsize
\newline
\begin{flushleft} \footnotesize \linespread{.75} Panel A compares two networks surveyed from 129 students in the ``Teenage Friends and Lifestyle Study.'' Panel B compares two networks from village 10 in \cite{banerjee2013diffusion}. Both panels describe point estimates and p-values (in parentheses) for the OLS regression of agent-level network centrality measures agent degree, eigenvector centrality, and agent clustering on a constant and an indicator for the second network. \normalsize \end{flushleft}
\end{table}

\small
\bibliographystyle{chicago}
\bibliography{literature}

\begin{thebibliography}{}

\bibitem[\protect\citeauthoryear{Alon and Naor}{Alon and
  Naor}{2006}]{alon2006approximating}
Alon, N. and A.~Naor (2006).
\newblock Approximating the cut-norm via grothendieck's inequality.
\newblock {\em SIAM Journal on Computing\/}~{\em 35\/}(4), 787--803.

\bibitem[\protect\citeauthoryear{Aronow}{Aronow}{2012}]{aronow2012general}
Aronow, P.~M. (2012).
\newblock A general method for detecting interference between units in
  randomized experiments.
\newblock {\em Sociological Methods \& Research\/}~{\em 41\/}(1), 3--16.

\bibitem[\protect\citeauthoryear{Athey, Eckles, and Imbens}{Athey
  et~al.}{2018}]{athey2018exact}
Athey, S., D.~Eckles, and G.~W. Imbens (2018).
\newblock Exact p-values for network interference.
\newblock {\em Journal of the American Statistical Association\/}~{\em
  113\/}(521), 230--240.

\bibitem[\protect\citeauthoryear{Bandeira and Van~Handel}{Bandeira and
  Van~Handel}{2016}]{bandeira2016sharp}
Bandeira, A.~S. and R.~Van~Handel (2016).
\newblock Sharp nonasymptotic bounds on the norm of random matrices with
  independent entries.
\newblock {\em The Annals of Probability\/}~{\em 44\/}(4), 2479--2506.

\bibitem[\protect\citeauthoryear{Banerjee, Chandrasekhar, Duflo, and
  Jackson}{Banerjee et~al.}{2013}]{banerjee2013diffusion}
Banerjee, A., A.~G. Chandrasekhar, E.~Duflo, and M.~O. Jackson (2013).
\newblock The diffusion of microfinance.
\newblock {\em Science\/}~{\em 341\/}(6144), 1236498.

\bibitem[\protect\citeauthoryear{Banerjee, Chandrasekhar, Duflo, and
  Jackson}{Banerjee et~al.}{2018}]{banerjee2018changes}
Banerjee, A.~V., A.~G. Chandrasekhar, E.~Duflo, and M.~O. Jackson (2018).
\newblock Changes in social network structure in response to exposure to formal
  credit markets.
\newblock {\em Available at SSRN 3245656\/}.

\bibitem[\protect\citeauthoryear{Boucheron, Lugosi, and Massart}{Boucheron
  et~al.}{2013}]{boucheron2013concentration}
Boucheron, S., G.~Lugosi, and P.~Massart (2013).
\newblock {\em Concentration inequalities: A nonasymptotic theory of
  independence}.
\newblock Oxford university press.

\bibitem[\protect\citeauthoryear{Calv{\'o}-Armengol, Patacchini, and
  Zenou}{Calv{\'o}-Armengol et~al.}{2009}]{calvo2009peer}
Calv{\'o}-Armengol, A., E.~Patacchini, and Y.~Zenou (2009).
\newblock Peer effects and social networks in education.
\newblock {\em The Review of Economic Studies\/}~{\em 76\/}(4), 1239--1267.

\bibitem[\protect\citeauthoryear{Fafchamps and Gubert}{Fafchamps and
  Gubert}{2007}]{fafchamps2007risk}
Fafchamps, M. and F.~Gubert (2007).
\newblock Risk sharing and network formation.
\newblock {\em American Economic Review\/}~{\em 97\/}(2), 75--79.

\bibitem[\protect\citeauthoryear{Ghoshdastidar, Gutzeit, Carpentier, and von
  Luxburg}{Ghoshdastidar et~al.}{2017a}]{ghoshdastidar2017two1}
Ghoshdastidar, D., M.~Gutzeit, A.~Carpentier, and U.~von Luxburg (2017a).
\newblock Two-sample hypothesis testing for inhomogeneous random graphs.
\newblock {\em arXiv preprint arXiv:1707.00833\/}.

\bibitem[\protect\citeauthoryear{Ghoshdastidar, Gutzeit, Carpentier, and von
  Luxburg}{Ghoshdastidar et~al.}{2017b}]{ghoshdastidar2017two2}
Ghoshdastidar, D., M.~Gutzeit, A.~Carpentier, and U.~von Luxburg (2017b).
\newblock Two-sample tests for large random graphs using network statistics.
\newblock {\em arXiv preprint arXiv:1705.06168\/}.

\bibitem[\protect\citeauthoryear{Gittens and Tropp}{Gittens and
  Tropp}{2009}]{gittens2009error}
Gittens, A. and J.~A. Tropp (2009).
\newblock Error bounds for random matrix approximation schemes.
\newblock {\em arXiv preprint arXiv:0911.4108\/}.

\bibitem[\protect\citeauthoryear{Goldsmith-Pinkham and
  Imbens}{Goldsmith-Pinkham and Imbens}{2013}]{goldsmith2013social}
Goldsmith-Pinkham, P. and G.~W. Imbens (2013).
\newblock Social networks and the identification of peer effects.
\newblock {\em Journal of Business \& Economic Statistics\/}~{\em 31\/}(3),
  253--264.

\bibitem[\protect\citeauthoryear{Goyal, Van Der~Leij, and
  Moraga-Gonz{\'a}lez}{Goyal et~al.}{2006}]{goyal2006economics}
Goyal, S., M.~J. Van Der~Leij, and J.~L. Moraga-Gonz{\'a}lez (2006).
\newblock Economics: An emerging small world.
\newblock {\em Journal of political economy\/}~{\em 114\/}(2), 403--412.

\bibitem[\protect\citeauthoryear{Graham}{Graham}{2019}]{graham2019network}
Graham, B.~S. (2019).
\newblock Network data.
\newblock {\em Forthcoming Econometrics Handbook Chapter\/}.

\bibitem[\protect\citeauthoryear{Imbens and Rubin}{Imbens and
  Rubin}{2015}]{imbensRubin2015}
Imbens, G.~W. and D.~B. Rubin (2015).
\newblock {\em Causal inference in statistics, social, and biomedical
  sciences}.
\newblock Cambridge University Press.

\bibitem[\protect\citeauthoryear{Jackson and Rogers}{Jackson and
  Rogers}{2007}]{jackson2007meeting}
Jackson, M.~O. and B.~W. Rogers (2007).
\newblock Meeting strangers and friends of friends: How random are social
  networks?
\newblock {\em American Economic Review\/}~{\em 97\/}(3), 890--915.

\bibitem[\protect\citeauthoryear{Jackson, Rogers, and Zenou}{Jackson
  et~al.}{2017}]{jackson2017economic}
Jackson, M.~O., B.~W. Rogers, and Y.~Zenou (2017).
\newblock The economic consequences of social-network structure.
\newblock {\em Journal of Economic Literature\/}~{\em 55\/}(1), 49--95.

\bibitem[\protect\citeauthoryear{Krivine}{Krivine}{1979}]{krivine1979constantes}
Krivine, J.-L. (1979).
\newblock Constantes de grothendieck et fonctions de type positif sur les
  spheres.
\newblock {\em Advances in Mathematics\/}~{\em 31\/}(1), 16--30.

\bibitem[\protect\citeauthoryear{Lehmann and Romano}{Lehmann and
  Romano}{2006}]{lehmann2006testing}
Lehmann, E.~L. and J.~P. Romano (2006).
\newblock {\em Testing statistical hypotheses}.
\newblock Springer Science \& Business Media.

\bibitem[\protect\citeauthoryear{Michell and West}{Michell and
  West}{1996}]{michell1996peer}
Michell, L. and P.~West (1996).
\newblock Peer pressure to smoke: the meaning depends on the method.
\newblock {\em Health education research\/}~{\em 11\/}(1), 39--49.

\bibitem[\protect\citeauthoryear{Nielsen and Witten}{Nielsen and
  Witten}{2018}]{nielsen2018multiple}
Nielsen, A.~M. and D.~Witten (2018).
\newblock The multiple random dot product graph model.
\newblock {\em arXiv preprint arXiv:1811.12172\/}.

\bibitem[\protect\citeauthoryear{Pelican and Graham}{Pelican and
  Graham}{2020}]{pelican2020optimal}
Pelican, A. and B.~S. Graham (2020).
\newblock An optimal test for strategic interaction in social and economic
  network formation between heterogeneous agents.
\newblock Technical report, National Bureau of Economic Research.

\bibitem[\protect\citeauthoryear{Rose}{Rose}{2004}]{rose2004we}
Rose, A.~K. (2004).
\newblock Do we really know that the wto increases trade?
\newblock {\em American economic review\/}~{\em 94\/}(1), 98--114.

\bibitem[\protect\citeauthoryear{Tang, Athreya, Sussman, Lyzinski, and
  Priebe}{Tang et~al.}{2017}]{tang2017nonparametric}
Tang, M., A.~Athreya, D.~L. Sussman, V.~Lyzinski, and C.~E. Priebe (2017).
\newblock A nonparametric two-sample hypothesis testing problem for random
  graphs.
\newblock {\em Bernoulli\/}~{\em 23\/}(3), 1599--1630.

\end{thebibliography}


\begin{thebibliography}{}

\bibitem[\protect\citeauthoryear{Arduini, Patacchini, and Rainone}{Arduini
  et~al.}{2015}]{arduini2015parametric}
Arduini, T., E.~Patacchini, and E.~Rainone (2015).
\newblock Parametric and semiparametric iv estimation of network models with
  selectivity.
\newblock Technical report, Einaudi Institute for Economics and Finance (EIEF).

\bibitem[\protect\citeauthoryear{Auerbach}{Auerbach}{2019}]{auerbach2019identification}
Auerbach, E. (2019).
\newblock Identification and estimation of a partially linear regression model
  using network data.
\newblock {\em arXiv preprint arXiv:1903.09679\/}.

\bibitem[\protect\citeauthoryear{Banerjee, Chandrasekhar, Duflo, and
  Jackson}{Banerjee et~al.}{2013}]{banerjee2013diffusion}
Banerjee, A., A.~G. Chandrasekhar, E.~Duflo, and M.~O. Jackson (2013).
\newblock The diffusion of microfinance.
\newblock {\em Science\/}~{\em 341\/}(6144), 1236498.

\bibitem[\protect\citeauthoryear{Banerjee, Chandrasekhar, Duflo, and
  Jackson}{Banerjee et~al.}{2018}]{banerjee2018changes}
Banerjee, A.~V., A.~G. Chandrasekhar, E.~Duflo, and M.~O. Jackson (2018).
\newblock Changes in social network structure in response to exposure to formal
  credit markets.
\newblock {\em Available at SSRN 3245656\/}.

\bibitem[\protect\citeauthoryear{Bloch and Jackson}{Bloch and
  Jackson}{2007}]{bloch2007formation}
Bloch, F. and M.~O. Jackson (2007).
\newblock The formation of networks with transfers among players.
\newblock {\em Journal of Economic Theory\/}~{\em 133\/}(1), 83--110.

\bibitem[\protect\citeauthoryear{Boucheron, Lugosi, and Massart}{Boucheron
  et~al.}{2013}]{boucheron2013concentration}
Boucheron, S., G.~Lugosi, and P.~Massart (2013).
\newblock {\em Concentration inequalities: A nonasymptotic theory of
  independence}.
\newblock Oxford university press.

\bibitem[\protect\citeauthoryear{Calv{\'o}-Armengol, Patacchini, and
  Zenou}{Calv{\'o}-Armengol et~al.}{2009}]{calvo2009peer}
Calv{\'o}-Armengol, A., E.~Patacchini, and Y.~Zenou (2009).
\newblock Peer effects and social networks in education.
\newblock {\em The Review of Economic Studies\/}~{\em 76\/}(4), 1239--1267.

\bibitem[\protect\citeauthoryear{Fafchamps and Gubert}{Fafchamps and
  Gubert}{2007}]{fafchamps2007risk}
Fafchamps, M. and F.~Gubert (2007).
\newblock Risk sharing and network formation.
\newblock {\em American Economic Review\/}~{\em 97\/}(2), 75--79.

\bibitem[\protect\citeauthoryear{Goldsmith-Pinkham and
  Imbens}{Goldsmith-Pinkham and Imbens}{2013}]{goldsmith2013social}
Goldsmith-Pinkham, P. and G.~W. Imbens (2013).
\newblock Social networks and the identification of peer effects.
\newblock {\em Journal of Business \& Economic Statistics\/}~{\em 31\/}(3),
  253--264.

\bibitem[\protect\citeauthoryear{Goyal, Van Der~Leij, and
  Moraga-Gonz{\'a}lez}{Goyal et~al.}{2006}]{goyal2006economics}
Goyal, S., M.~J. Van Der~Leij, and J.~L. Moraga-Gonz{\'a}lez (2006).
\newblock Economics: An emerging small world.
\newblock {\em Journal of political economy\/}~{\em 114\/}(2), 403--412.

\bibitem[\protect\citeauthoryear{Graham}{Graham}{2015}]{graham2015methods}
Graham, B.~S. (2015).
\newblock Methods of identification in social networks.
\newblock {\em Annu. Rev. Econ.\/}~{\em 7\/}(1), 465--485.

\bibitem[\protect\citeauthoryear{Hsieh and Lee}{Hsieh and
  Lee}{2014}]{hsieh2014social}
Hsieh, C.-S. and L.~F. Lee (2014).
\newblock A social interactions model with endogenous friendship formation and
  selectivity.
\newblock {\em Journal of Applied Econometrics\/}.

\bibitem[\protect\citeauthoryear{Imbens and Rubin}{Imbens and
  Rubin}{2015}]{imbensRubin2015}
Imbens, G.~W. and D.~B. Rubin (2015).
\newblock {\em Causal inference in statistics, social, and biomedical
  sciences}.
\newblock Cambridge University Press.

\bibitem[\protect\citeauthoryear{Jackson and Rogers}{Jackson and
  Rogers}{2007}]{jackson2007meeting}
Jackson, M.~O. and B.~W. Rogers (2007).
\newblock Meeting strangers and friends of friends: How random are social
  networks?
\newblock {\em American Economic Review\/}~{\em 97\/}(3), 890--915.

\bibitem[\protect\citeauthoryear{Jackson, Rogers, and Zenou}{Jackson
  et~al.}{2017}]{jackson2017economic}
Jackson, M.~O., B.~W. Rogers, and Y.~Zenou (2017).
\newblock The economic consequences of social-network structure.
\newblock {\em Journal of Economic Literature\/}~{\em 55\/}(1), 49--95.

\bibitem[\protect\citeauthoryear{Johnsson and Moon}{Johnsson and
  Moon}{2015}]{johnsson2015estimation}
Johnsson, I. and H.~R. Moon (2015).
\newblock Estimation of peer effects in endogenous social networks: Control
  function approach.

\bibitem[\protect\citeauthoryear{Pelican and Graham}{Pelican and
  Graham}{2020}]{pelican2020optimal}
Pelican, A. and B.~S. Graham (2020).
\newblock An optimal test for strategic interaction in social and economic
  network formation between heterogeneous agents.
\newblock Technical report, National Bureau of Economic Research.

\bibitem[\protect\citeauthoryear{Rose}{Rose}{2004}]{rose2004we}
Rose, A.~K. (2004).
\newblock Do we really know that the wto increases trade?
\newblock {\em American economic review\/}~{\em 94\/}(1), 98--114.

\end{thebibliography}
\footnotesize
\appendix

\section{Proofs}
\subsection{Theorem 1}
\begin{flushleft}
\textbf{Lemma 1}: Let $X$ be an arbitrary $N\times N$ dimensional random symmetric matrix with independent and mean-zero entries above the diagonal and zeros on the main diagonal. The entries of $X$ are absolutely bounded by $1$. Let $\{X_{s}\}_{s \in S}$ be an arbitrary collection of such matrices of size $S$. Then for any fixed $\alpha \in [0,1]$ and $\gamma \in [0,1/2]$
\begin{align*}
\max_{s \in [S]}\max_{j \in [N]} \sqrt{\sum_{i\in[N]}X_{ij,s}^{2}} \leq \max_{s \in [S]}\max_{\varphi \in \mathcal{S}^{N}}\sqrt{\sum_{i \in [N]}\left(\sum_{j \in [N]}X_{ij,s}\varphi_{j}\right)^{2}} \\
\leq \sqrt{-2\ln\left(\frac{\alpha}{S}\right)} +  (1+ \gamma)2\max_{s \in [S]}\max_{j \in [N]}\sqrt{\sum_{i\in[N]}E\left[X_{ij,s}^{2}\right]} + \frac{6(1+\gamma)}{\sqrt{\ln(1+\gamma)}}\sqrt{\ln (N)}
\end{align*}
 with probability at least $1-\alpha$. $\square$
\end{flushleft} 
\vspace{2mm}
\begin{flushleft}
\textbf{Proof of Lemma 1}:  The lower bound holds for any collection of matrices. It follows from
\begin{align*}
\max_{\varphi \in \mathcal{S}^{N}}\sum_{i \in [N]}\left(\sum_{j \in [N]}X_{ij}\varphi_{j}\right)^{2} \geq\max_{\varphi \in \mathcal{E}^{N}}\sum_{i \in [N]}\left(\sum_{j \in [N]}X_{ij}\varphi_{j}\right)^{2}  = \max_{j \in [N]} \sum_{i\in[N]}X_{ij}^{2}
\end{align*}
where $ \mathcal{S}^{N}$ is the $N$-dimensional hypersphere $\{\varphi \in \mathbb{R}^{N}: \sum_{t \in [N]}\varphi_{t}^{2} = 1\}$, $\mathcal{E}^{N}$ is the usual set of basis vectors in $\mathbb{R}^{N}$ $\{\varphi \in \mathbb{R}^{N}: \sum_{t\in[N]}\varphi_{t}^{2} = \sum_{t \in [N]}\left|\varphi_{j}\right| = 1\}$, and the inequality follows from $\mathcal{E}^{N} \subset \mathcal{S}^{N}$. Consequently, if $\{X_{s}\}_{s \in S}$ is any collection of $N\times N$ dimensional matrices indexed by a set $S$ then
\begin{align*}
\max_{s \in [S]}\max_{\varphi \in \mathcal{S}^{N}}\sqrt{\sum_{i \in [N]}\left(\sum_{j \in [N]}X_{ij,s}\varphi_{j}\right)^{2}} \geq  \max_{s \in [S]}\max_{j \in [N]} \sqrt{\sum_{i\in[N]}X_{ij,s}^{2}}
\end{align*} 
where $X_{ij,s}$ is the $ij$th entry of the matrix $X_{s}$.\newline

The upper bound follows inequalities by Talagrand \citep*[see][Theorem 6.10]{boucheron2013concentration} and \cite*{bandeira2016sharp}. Specifically, for any  $\varepsilon > 0$, Talagrand's inequality implies 
\begin{align*}
P\left(\max_{\varphi \in \mathcal{S}^{N}}\sqrt{\sum_{i \in [N]}\left(\sum_{j \in [N]}X_{ij}\varphi_{j}\right)^{2}}- E\left[\max_{\varphi \in \mathcal{S}^{N}}\sqrt{\sum_{i \in [N]}\left(\sum_{j \in [N]}X_{ij}\varphi_{j}\right)^{2}}\right] > \varepsilon\right) \leq \exp\left(-\varepsilon^{2}/2\right).
\end{align*}
since $\max_{\varphi \in \mathcal{S}^{N}}\sqrt{\sum_{i \in [N]}\left(\sum_{j \in [N]}X_{ij}\varphi_{j}\right)^{2}}$ is convex in $X$ by the triangle inequality. \newline
 
Corollary 3.2 to Theorem 1.1 of \cite*{bandeira2016sharp} implies 
\begin{align*}
E\left[\max_{\varphi \in \mathcal{S}^{N}}\sqrt{\sum_{i \in [N]}\left(\sum_{j \in [N]}X_{ij}\varphi_{j}\right)^{2}}\right] \leq  (1+\gamma)\left[2\max_{j \in [N]} \sqrt{\sum_{i\in[N]}E\left[X_{ij}^{2}\right]} + \frac{6}{\sqrt{\ln(1+\gamma)}}\sqrt{\ln (N)}\right]
\end{align*}
 for any $\gamma \in [0,1/2]$. Consequently, for any real positive integer $S$, collection of $N\times N$ dimensional random symmetric matrices $\{X_{s}\}_{s \in [S]}$ such that each matrix $X_{s}$ satisfies the above conditions, and $\gamma \in [0,1/2]$
\begin{align*}
\max_{s \in [S]}\max_{\varphi \in \mathcal{S}^{N}}\sqrt{\sum_{i \in [N]}\left(\sum_{j \in [N]}X_{ij}\varphi_{j}\right)^{2}} > \varepsilon +  (1+ \gamma)2\max_{s \in [S]}\max_{j \in [N]}\sqrt{\sum_{i\in[N]}E\left[X_{ij,s}^{2}\right]} + \frac{6(1+\gamma)}{\sqrt{\ln(1+\gamma)}}\sqrt{\ln (N)}
\end{align*}
with probability less than $S\exp\left(-\varepsilon^{2}/2\right)$ by the union bound. Or equivalently, for any $\alpha \in [0,1]$ and $\gamma \in [0,1/2]$
\begin{align*}
\max_{s \in [S]}\max_{\varphi \in \mathcal{S}^{N}}\sqrt{\sum_{i \in [N]}\left(\sum_{j \in [N]}X_{ij}\varphi_{j}\right)^{2}} \leq \sqrt{-2\ln\left(\frac{\alpha}{S}\right)} +  (1+ \gamma)2\max_{s \in [S]}\max_{j \in [N]}\sqrt{\sum_{i\in[N]}E\left[X_{ij,s}^{2}\right]} + \frac{6(1+\gamma)}{\sqrt{\ln(1+\gamma)}}\sqrt{\ln (N)}
\end{align*}
with probability greater than $1-\alpha$. The claim follows. $\square$
\end{flushleft}

\begin{flushleft}
\textbf{Proof of Theorem 1}: The main part of the proof is to show that for any $r \in [R]$,
$T_{2\to2}(D_{1},D_{2}) \geq T_{2\to2}(D_{1}^{r},D_{2}^{r})$ with high probability. The first step is to bound $T_{2\to2}(D_{1}^{r},D_{2}^{r}) = \max_{s \in \mathbb{R}}\max_{\varphi \in \mathcal{S}^{N}}\sqrt{\sum_{i \in [N]}\left(\sum_{j \in [N]}\left(\mathbbm{1}_{D_{ij,1}^{r} \leq s} - \mathbbm{1}_{D_{ij,2}^{r} \leq s}\right)\varphi_{j}\right)^{2}}$ from above. Since for any $r \in [R]$, $\{\mathbbm{1}_{D_{1}^{r} \leq s} - \mathbbm{1}_{D_{2}^{r} \leq s}\}_{s \in \mathbb{R}}$ is a collection of no more than $N(N-1)$ unique $N\times N$ dimensional random symmetric matrices with independent and mean-zero entries above the diagonal and zeros on the main diagonal, all absolutely bounded by $1$, Lemma 1 implies that for any $\beta \in [0,1]$ and $\gamma \in [0,1/2]$
\begin{align*}
\max_{s \in \mathbb{R}}&\max_{\varphi \in \mathcal{S}^{N}}\sqrt{\sum_{i \in [N]}\left(\sum_{j \in [N]}\left(\mathbbm{1}_{D_{ij,1}^{r} \leq s} - \mathbbm{1}_{D_{ij,2}^{r} \leq s}\right)\varphi_{j}\right)^{2}} 
\leq \sqrt{-2\ln\left(\frac{\beta}{N(N-1)}\right)} \\
&+  (1+ \gamma)2\max_{s \in [\mathbb{R}]}\max_{j \in [N]}\sqrt{\sum_{i\in[N]}\nu_{ij}(s)}  + \frac{6(1+\gamma)}{\sqrt{\ln(1+\gamma)}}\sqrt{\ln (N)}
\end{align*}
with probability greater than $1-\beta$, where $\nu_{ij}(s) = \left[F_{ij,1}(s) + F_{ij,2}(s) - 2F_{ij,1}(s)F_{ij,2}(s)\right]$. Since $-\ln(\alpha)/\ln(N) = O(1)$ by assumption in Section 4.2.1, it is possible to choose $\beta$ such that $-\ln(\beta)/\ln(N) = O(1)$ and $\beta = o(\alpha)$. Along with the rate condition $\tau/\sqrt{\ln(N)} \to \infty$, this implies 
\begin{align*}
 \frac{\sqrt{-\ln(\beta) + \ln(N)}}{\tau} \to 0
\end{align*}
and so
\begin{align*}
\frac{\max_{s \in \mathbb{R}}\max_{\varphi \in \mathcal{S}^{N}}\sqrt{\sum_{i \in [N]}\left(\sum_{j \in [N]}\left(\mathbbm{1}_{D_{ij,1}^{r} \leq s} - \mathbbm{1}_{D_{ij,2}^{r} \leq s}\right)\varphi_{j}\right)^{2}}}{\tau} \leq 3
\end{align*} 
with probability greater than $1- \beta$ eventually. 
\newline

The second step is to bound $T_{2\to2}(D_{1},D_{2}) = \max_{s \in \mathbb{R}}\max_{\varphi \in \mathcal{S}^{N}}\sqrt{\sum_{i \in [N]}\left(\sum_{j \in [N]}\left(\mathbbm{1}_{D_{ij,1} \leq s} - \mathbbm{1}_{D_{ij,2} 
\leq s}\right)\varphi_{j}\right)^{2}}$ from below. By two applications of the triangle inequality, 
\begin{align*}
\max_{s \in \mathbb{R}}&\max_{\varphi \in \mathcal{S}^{N}}\sqrt{\sum_{i \in [N]}\left(\sum_{j \in [N]}\left(\mathbbm{1}_{D_{ij,1} \leq s} - \mathbbm{1}_{D_{ij,2} 
\leq s}\right)\varphi_{j}\right)^{2}} 
\geq \max_{s \in \mathbb{R}}\max_{\varphi \in \mathcal{S}^{N}}\sqrt{\sum_{i \in [N]}\left(\sum_{j \in [N]}\left(F_{ij,1}(s) -F_{ij,2}(s)\right)\varphi_{j}\right)^{2}} \\
&- \left|\max_{s \in \mathbb{R}}\max_{\varphi \in \mathcal{S}^{N}}\sqrt{\sum_{i \in [N]}\left(\sum_{j \in [N]}\left(\mathbbm{1}_{D_{ij,1}\leq s} -F_{ij,1}(s)\right)\varphi_{j}\right)^{2}} + \max_{s \in \mathbb{R}}\max_{\varphi \in \mathcal{S}^{N}}\sqrt{\sum_{i \in [N]}\left(\sum_{j \in [N]}\left(\mathbbm{1}_{D_{ij,2} \leq s} -F_{ij,2}(s)\right)\varphi_{j}\right)^{2}}\right|. 
\end{align*}

The matrices $\{\mathbbm{1}_{D_{t}\leq s} -F_{t}(s)\}_{s \in \mathbb{R}, t \in \{1,2\}}$ also satisfy the hypothesis of Lemma 1, and so too
\begin{align*}
\frac{\max_{s \in \mathbb{R}}\max_{\varphi \in \mathcal{S}^{N}}\sqrt{\sum_{i \in [N]}\left(\sum_{j \in [N]}\left(\mathbbm{1}_{D_{ij,t} \leq s}  - F_{ij,t}(s)\right)\varphi_{j}\right)^{2}}}{\tau}  \leq 3.
\end{align*}
for $t \in [2]$ and with probability greater than $(1-2\beta)$ eventually. The rate condition $T_{2\to2}(F_{1},F_{2})/\tau \to \infty$ then implies 
\begin{align*}
\frac{\max_{s \in \mathbb{R}}\max_{\varphi \in \mathcal{S}^{N}}\sqrt{\sum_{i \in [N]}\left(\sum_{j \in [N]}\left(F_{ij,1}(s) -F_{ij,2}(s)\right)\varphi_{j}\right)^{2}}}{\tau} \to \infty,
\end{align*}
and so 
\begin{align*}
\frac{\max_{s \in \mathbb{R}}\max_{\varphi \in \mathcal{S}^{N}}\sqrt{\sum_{i \in [N]}\left(\sum_{j \in [N]}\left(\mathbbm{1}_{D_{ij,1} \leq s} - \mathbbm{1}_{D_{ij,2}\leq s}\right)\varphi_{j}\right)^{2}}}{\tau} \to \infty.
\end{align*}
This demonstrates that $T_{2\to2}(D_{1},D_{2}) \geq T_{2\to2}(D_{1}^{r},D_{2}^{r})$  with probability greater than $1-3\beta$ for any $r \in [R]$. 
\newline

To finish the proof, write 
\begin{align*}
P\left((R+1)^{-1}\left(1 + \sum_{r\in[R]}\mathbbm{1}\{T_{\infty\to1}(D_{1}^{r},D_{2}^{r}) \geq T_{2\to2}(D_{1},D_{2})\} \right) \leq \alpha\right) \\
= P\left(R^{-1}\sum_{r\in[R]}\mathbbm{1}\{T_{2\to2}(D_{1}^{r},D_{2}^{r}) \geq T_{2\to2}(D_{1},D_{2})\}  \leq \frac{\alpha(R+1) - 1}{R}\right).
\end{align*}
On the right-hand side, $\frac{\alpha(R+1) - 1}{R} \geq \alpha/2$ since $\alpha R \geq 2$ by assumption in Section 4.2.1. On the left-hand side, the entries of $\{\mathbbm{1}\{T_{2\to2}(D_{1}^{r},D_{2}^{r}) \geq T_{2\to2}(D_{1},D_{2})\}\}_{r \in [R]}$ are independent and $o_{p}(\alpha)$, and so their average is $o_{p}(\alpha)$. The claim follows.  $\square$.
\end{flushleft}

\subsection{Theorem 2}
The results below rely on the following inequality due to Grothendieck and \cite*{krivine1979constantes}, see generally \cite{alon2006approximating}.
\begin{flushleft}
\textbf{Theorem (Grothendieck):} Let $X$ be an arbitrary $N\times N$ dimensional real matrix such that 
\begin{align*}
\max_{\varphi,\psi \in \mathbb{R}^{N}: ||\varphi||_{\infty},||\psi||_{\infty} \leq 1}\left|\sum_{i \in [N]}\sum_{j \in [N]}X_{ij}\varphi_{j}\psi_{i}\right| \leq 1.
\end{align*}
Then 
\begin{align*}
\max_{\varphi,\psi \in \mathcal{H}: ||\varphi||_{\mathcal{H}},||\psi||_{\mathcal{H}} \leq 1}\left|\sum_{i \in [N]}\sum_{j \in [N]}X_{ij}<\tilde{\varphi}_{i},\tilde{\psi}_{j}>_{\mathcal{H}}\right| \leq K = \frac{\pi}{2\ln\left(1 + \sqrt{2}\right)} \leq 1.783
\end{align*} 
where $\mathcal{H}$ is an arbitrary Hilbert space and $||\cdot||_{\mathcal{H}}$ and $<\cdot,\cdot>_{\mathcal{H}}$ are the associated norm and inner product operators. 
\end{flushleft}

\begin{flushleft}
\textbf{Bounds from Section 4.1:} For any $D_{1}, D_{0}$ and $N$, 
\begin{align*}
T_{\infty\to1} \leq S_{\infty\to1} \leq \frac{\pi}{2\ln{\left(1 + \sqrt{2}\right)}}T_{\infty\to1} \leq 1.783\hspace{1mm} T_{\infty\to1}.
\end{align*}
\end{flushleft}
\vspace{2mm}

\begin{flushleft}
\textbf{Proof of claim}:
The first inequality follows from 
 \begin{align*}
T_{\infty\to1} &=  \max_{s \in \mathbb{R}}\max_{\varphi : ||\varphi||_{\infty} = 1}||\Delta(s)\varphi||_{1} = \max_{s \in \mathbb{R}}\max_{\varphi: ||\varphi||_{\infty} = 1}\max_{\psi: ||\psi||_{\infty} = 1}\left<\Delta(s),\varphi\otimes\psi\right> \\
&= \frac{1}{2}\max_{s \in \mathbb{R}}\max_{\phi:  ||\phi||_{\infty} = 1}\left<\begin{bmatrix} 0_{N\times N} & \Delta(s) \\ \Delta(s) & 0_{N\times N}  \end{bmatrix},\phi\otimes\phi\right> \leq \frac{1}{2}\max_{s \in \mathbb{R}}\max_{X \in \mathcal{X}_{2N}}\left<\begin{bmatrix} 0_{N\times N} & \Delta(s) \\ \Delta(s) & 0_{N\times N}  \end{bmatrix},X\right>  = S_{\infty\to1}
\end{align*} 
where $\otimes$ refers to the vector outer product operator, the first equality follows from taking $\psi = \text{sign}\left(\Delta(s)\varphi\right)$, the second equality follows by choosing $\phi$ to be the concatenation of $\varphi$ and $\psi$, and the inequality follows from $\phi \otimes \phi  \in \mathcal{X}_{2N}$. Grothendieck's inequality directly implies the second inequality. 
\end{flushleft}
\vspace{2mm}

\begin{flushleft}
\textbf{Lemma 2}: Let $X$ be an arbitrary $N\times N$ dimensional random symmetric matrix with independent and mean-zero entries above the diagonal and zeros on the main diagonal. The entries of $X$ are absolutely bounded by $1$. Let $\{X_{s}\}_{s \in S}$ be an arbitrary collection of such matrices of size $S$. Then for any fixed $\alpha \in [0,1]$
\begin{align*}
\frac{1}{K}\max_{s \in [S]} \sum_{i\in [N]}\sqrt{\sum_{j\in [N]}X_{ij,s}^{2}}   \leq \max_{s \in [S]}\max_{\varphi \in \mathcal{C}^{N}}\sum_{i \in [N]}\left|\sum_{j \in [N]}X_{ij,s}\varphi_{j}\right| \\
\leq \sqrt{-2\ln\left(\frac{\alpha}{S}\right)} + 2\max_{s \in [S]}\sum_{i\in [N]}\sqrt{\sum_{j\in [N]}E\left[X_{ij,s}^{2}\right]} 
\end{align*}
 with probability at least $1-\alpha$ where $K = 1.783$ is Grothendieck's constant. $\square$
 \end{flushleft} 
\vspace{2mm}

\begin{flushleft}
\textbf{Proof of  Lemma 2}: The lower bound holds for any collection of matrices. It follows from
\begin{align*}
 &K\max_{\varphi \in \mathcal{C}^{N}}\sum_{i \in [N]}\left|\sum_{j \in [N]}X_{ij}\varphi_{j}\right| = K\max_{\varphi,\psi \in \mathcal{C}^{N}}\sum_{i \in [N]}\sum_{j \in [N]}X_{ij}\varphi_{j}\psi_{i} 
 \geq  \max_{\varphi,\psi \in \mathcal{M}^{N}}\sum_{i \in [N]}\sum_{j \in [N]}X_{ij}\sum_{s \in [N]}\varphi_{js}\psi_{is} \\
 &\geq \sum_{i \in [N]}\sum_{j\in[N]}X_{ij}\sum_{s\in [N]}\frac{X_{is}}{\sqrt{\sum_{s\in [N]}X_{is}^{2}}}\mathbbm{1}\{j=s\} = \sum_{i\in [N]}\sqrt{\sum_{j\in [N]}X_{ij}^{2}} 
\end{align*}
where $\mathcal{C}^{N}$ is the $N$-dimensional hypercube $ \{-1,1\}^{N}$, $\mathcal{M}^{N}$ is the set of $N\times N$ matrices with rows of Euclidean length 1 $\{\Lambda \in \mathbb{R}^{N\times N}: \sum_{j=1}^{N}\Lambda_{ij}^{2} = 1 \forall i \in [N]\}$, and the first inequality is due to Grothendieck. Consequently, if $\{X_{s}\}_{s \in S}$ is any collection of $N\times N$ dimensional matrices indexed by a set $S$ then
\begin{align*}
K\max_{s \in [S]}\max_{\varphi,\psi \in \mathcal{C}^{N}}\sum_{i \in [N]}\sum_{j \in [N]}X_{ij,s}\varphi_{j}\psi_{i}  \geq  \max_{s \in [S]} \sum_{i\in [N]}\sqrt{\sum_{j\in [N]}X_{ij,s}^{2}} 
\end{align*} 
where $X_{ij,s}$ is the $ij$th entry of matrix $X_{s}$. \newline

The upper bound also follows from Talagrand's inequality and an inequality due to \cite*{gittens2009error}. Specifically, let $X$ be an $N\times N$ dimensional random symmetric matrix with independent and mean-zero entries above the diagonal and zeros on the main diagonal. The entries of $X$ are absolutely bounded by 1. Then for any  $\varepsilon > 0$, Talagrand's inequality implies 
\begin{align*}
P\left(\max_{\varphi \in \mathcal{C}^{N}}\sum_{i \in [N]}\left|\sum_{j \in [N]}X_{ij}\varphi_{j}\right|- E\left[\max_{\varphi \in \mathcal{C}^{N}}\sum_{i \in [N]}\left|\sum_{j \in [N]}X_{ij}\varphi_{j}\right|\right] > \varepsilon\right) \leq \exp\left(-\varepsilon^{2}/2\right).
\end{align*}
since $\max_{\varphi \in \mathcal{C}^{N}}\sum_{i \in [N]}\left|\sum_{j \in [N]}X_{ij}\varphi_{j}\right|$ is convex in $X$ by the triangle inequality. \newline

Corollary 2  to Theorem 3 of \cite*{gittens2009error} implies 
\begin{align*}
E\left[\max_{\varphi \in \mathcal{C}^{N}}\sum_{i \in [N]}\left|\sum_{j \in [N]}X_{ij}\varphi_{j}\right|\right] \leq  2\sum_{i\in [N]}\sqrt{\sum_{j\in [N]}E\left[X_{ij}^{2}\right]}.
\end{align*}
Consequently, for any real positive integer $S$ and collection of $N\times N$ dimensional random symmetric matrices $\{X_{s}\}_{s \in [S]}$ such that each matrix $X_{s}$ satisfies the above conditions
\begin{align*}
\max_{s \in [S]}\max_{\varphi \in \mathcal{C}^{N}}\sum_{i \in [N]}\left|\sum_{j \in [N]}X_{ij,s}\varphi_{j}\right| > \varepsilon + 2\max_{s \in [S]}\sum_{i\in [N]}\sqrt{\sum_{j\in [N]}E\left[X_{ij,s}^{2}\right]} 
\end{align*}
with probability less than $S\exp\left(-\varepsilon^{2}/2\right)$ by the union bound. So for any $\alpha \in [0,1]$
\begin{align*}
\max_{s \in [S]}\max_{\varphi \in \mathcal{C}^{N}}\sum_{i \in [N]}\left|\sum_{j \in [N]}X_{ij,s}\varphi_{j}\right| \leq \sqrt{-2\ln\left(\frac{\alpha}{S}\right)} + 2\max_{s \in [S]}\sum_{i\in [N]}\sqrt{\sum_{j\in [N]}E\left[X_{ij,s}^{2}\right]} 
\end{align*}
with probability greater than $1-\alpha$. The claim follows. $\square$
\end{flushleft}

\begin{flushleft}
\textbf{Proof of Theorem 2}: The main part of the proof is to show that for any $r \in [R]$, $S_{\infty\to1}(D_{1},D_{2}) \geq S_{\infty\to1}(D_{1}^{r},D_{2}^{r})$ with high probability. The first step is to bound  $S_{\infty\to1}(D_{1}^{r},D_{2}^{r})  \leq KT_{\infty\to1}(D_{1}^{r},D_{2}^{r}) = K\max_{s \in \mathbb{R}}\max_{\varphi \in \mathcal{C}^{N}}\sum_{i \in [N]}\left|\sum_{j \in [N]}\left(\mathbbm{1}_{D_{ij,1}^{r} \leq s} - \mathbbm{1}_{D_{ij,2}^{r} \leq s}\right)\varphi_{j}\right|$ from above. Since for any $r \in [R]$, $\{\mathbbm{1}_{D_{1}^{r} \leq s} - \mathbbm{1}_{D_{2}^{r} \leq s}\}_{s \in \mathbb{R}}$ is a collection of no more than $N(N-1)$ unique $N\times N$ dimensional random symmetric matrices with independent and mean-zero entries above the diagonal and zeros on the main diagonal, all absolutely bounded by 1, Lemma 2 implies that for any $\beta \in [0,1]$
\begin{align*}
\max_{s \in \mathbb{R}}&\max_{\varphi \in \mathcal{C}^{N}}\sum_{i \in [N]}\left|\sum_{j \in [N]}\left(\mathbbm{1}_{D_{ij,1}^{r} \leq s} - \mathbbm{1}_{D_{ij,2}^{r} \leq s}\right)\varphi_{j}\right| 
\leq \sqrt{-2\ln\left(\frac{\beta}{N(N-1)}\right)} + 2\max_{s \in \mathbb{R}}\sum_{i\in [N]}\sqrt{\sum_{j\in [N]}\nu_{ij}(s)}
\end{align*}
with probability greater than $1-\beta$ where $\nu_{ij}(s) = \left[F_{ij,1}(s) + F_{ij,2}(s) - 2F_{ij,1}(s)F_{ij,2}(s)\right]$. Since $-\ln(\alpha)/\ln(N) = O(1)$ by assumption in Section 4.2.1, it is possible to choose $\beta$ such that $-\ln(\beta)/\ln(N) = O(1)$ and $\beta = o(\alpha)$. Along with the rate condition $\sigma/\sqrt{\ln(N)} \to \infty$, this implies
\begin{align*}
\frac{\sqrt{-\ln(\beta) + \ln(N)}}{\sigma} \to 0
\end{align*}
and so 
\begin{align*}
\frac{\max_{s \in \mathbb{R}}\max_{\varphi \in \mathcal{C}^{N}}\sum_{i \in [N]}\left|\sum_{j \in [N]}\left(\mathbbm{1}_{D_{ij,1}^{r}\leq s} -\mathbbm{1}_{D_{ij,2}^{r}\leq s}\right)\varphi_{j}\right|}{\sigma}  \leq 3
\end{align*}
with probability greater than $1 - \beta$ eventually. 
\newline

The second step is to bound $S_{\infty\to1}(D_{1},D_{2}) \geq \max_{s \in \mathbb{R}}\max_{\varphi \in \mathcal{C}^{N}}\sum_{i \in [N]}\left|\sum_{j \in [N]}\left(\mathbbm{1}_{D_{ij,1} \leq s} - \mathbbm{1}_{D_{ij,2} 
\leq s}\right)\varphi_{j}\right| $ from below. By two applications of the triangle inequality, 
\begin{align*}
\max_{s \in \mathbb{R}}&\max_{\varphi \in \mathcal{C}^{N}}\sum_{i \in [N]}\left|\sum_{j \in [N]}\left(\mathbbm{1}_{D_{ij,1} \leq s} - \mathbbm{1}_{D_{ij,2} 
\leq s}\right)\varphi_{j}\right| 
\geq \max_{s \in \mathbb{R}}\max_{\varphi \in \mathcal{C}^{N}}\sum_{i \in [N]}\left|\sum_{j \in [N]}\left(F_{ij,1}(s) -F_{ij,2}(s)\right)\varphi_{j}\right|\\
&- \left|\max_{s \in \mathbb{R}}\max_{\varphi \in \mathcal{C}^{N}}\sum_{i \in [N]}\left|\sum_{j \in [N]}\left(\mathbbm{1}_{D_{ij,1} \leq s} -F_{ij,1}(s)\right)\varphi_{j}\right| + \max_{s \in \mathbb{R}}\max_{\varphi \in \mathcal{C}^{N}}\sum_{i \in [N]}\left|\sum_{j \in [N]}\left(\mathbbm{1}_{D_{ij,2} \leq s} -F_{ij,2}(s)\right)\varphi_{j}\right|\right|. 
\end{align*}
The matrices $\{\mathbbm{1}_{D_{t} \leq s} - F_{t}(s)\}_{s \in \mathbb{R},t \in \{1,2\}}$ also satisfy the hypothesis of Lemma 2, and so too
\begin{align*}
\frac{\max_{s \in \mathbb{R}}\max_{\varphi \in \mathcal{C}^{N}}\sum_{i \in [N]}\left|\sum_{j \in [N]}\left(\mathbbm{1}_{D_{ij,t}\leq s} -F_{ij,t}(s)\right)\varphi_{j}\right|}{\sigma} \leq 3
\end{align*}
for $t \in [2]$ and with probability greater than $(1-2\beta)$ eventually. The rate condition $T_{\infty\to1}(F_{1},F_{2})/\sigma \to \infty$ implies
\begin{align*}
\frac{\max_{s \in \mathbb{R}}\max_{\varphi \in \mathcal{C}^{N}}\sum_{i \in [N]}\left|\sum_{j \in [N]}\left(F_{ij,1}(s) -F_{ij,2}(s)\right)\varphi_{j}\right|}{\sigma} \to \infty,
\end{align*}
and so
\begin{align*}
\frac{\max_{s \in \mathbb{R}}\max_{\varphi \in \mathcal{C}^{N}}\sum_{i \in [N]}\left|\sum_{j \in [N]}\left(\mathbbm{1}_{D_{ij,1}\leq s}-\mathbbm{1}_{D_{ij,2}\leq s}\right)\varphi_{j}\right|}{\sigma} \to \infty.
\end{align*}
This demonstrates that $S_{\infty\to1}(D_{1},D_{2}) \geq S_{\infty\to1}(D_{1}^{r},D_{2}^{r})$ with probability greater than $1-3\beta$ eventually for any $r \in [R]$.  
\newline

To finish the proof, write
\begin{align*}
P\left((R+1)^{-1}\left(1 + \sum_{r\in[R]}\mathbbm{1}\{S_{\infty\to1}(D_{1}^{r},D_{2}^{r}) \geq S_{\infty\to1}(D_{1},D_{2})\} \right) \leq \alpha\right) \\
= P\left(R^{-1}\sum_{r\in[R]}\mathbbm{1}\{S_{\infty\to1}(D_{1}^{r},D_{2}^{r}) \geq S_{\infty\to1}(D_{1},D_{2})\}  \leq \frac{\alpha(R+1) - 1}{R}\right).
\end{align*}
On the right-hand side $\frac{\alpha(R+1) - 1}{R} \geq \alpha/2$ by assumption in Section 4.2.1. On the left-hand side, the entries of $\{\mathbbm{1}\{S_{\infty\to1}(D_{1}^{r},D_{2}^{r}) \geq S_{\infty\to1}(D_{1},D_{2})\}\}_{r \in [R]}$ are independent and $o_{p}(\alpha)$, and so their average is $o_{p}(\alpha)$. The claim follows.  $\square$ 
\end{flushleft}

\end{document}


\maketitle

\footnotesize
\appendix
\stepcounter{section}

Section B  contains additional results about the power properties of the tests based on $T_{2\to2}$ and $S_{\infty\to1}$ (see Theorems 1 and 2 respectively in Section 4.2 of the main text). Section C contains results from Monte Carlo experiments. Section D contains details about the empirical applications and extensions described in Section 2.3 of the main text.  
\section{Additional Results}
\subsection{Near-Optimality of Rate Conditions}
The rate conditions  $T_{2\to2}(F_{1},F_{2})/\tau \to \infty $ and $T_{\infty\to1}(F_{1},F_{2})/\sigma \to \infty$ are close to necessary. 
\vspace{2mm}
\begin{flushleft}
\textbf{Theorem 3}: 
For any sequence of positive real numbers $\delta_{N} \to \infty$
\begin{align*}
\inf_{\substack{F_{1},F_{2}: \delta_{N} \left[T_{2\to2}(F_{1},F_{2})/\tau\right] \to \infty \\ \text{ and } \tau/\sqrt{\ln(N)} \to \infty}}
P\left((R+1)^{-1}\left(1 + \sum_{r\in[R]}\mathbbm{1}\{T_{2\to2}(D_{1}^{r},D_{2}^{r}) \geq T_{2\to2}(D_{1},D_{2})\} \right) \leq \alpha\right) \to \alpha
\end{align*}
and
\begin{align*}
\inf_{\substack{F_{1},F_{2}: \delta_{N} \left[T_{\infty\to1}(F_{1},F_{2})/\sigma\right] \to \infty \\ \text{ and } \sigma/\sqrt{\ln(N)} \to \infty}}
P\left((R+1)^{-1}\left(1 + \sum_{r\in[R]}\mathbbm{1}\{S_{\infty\to1}(D_{1}^{r},D_{2}^{r}) \geq S_{\infty\to1}(D_{1},D_{2})\} \right) \leq \alpha\right) \to \alpha.\hspace{5mm} \square
\end{align*}
\end{flushleft} 
\vspace{2mm}

The statement of Theorem 3 differs from that of Theorems 1 and 2 in two ways. The first is that the rate conditions in the subscript under the infima have been changed from $\left[T_{2\to2}(F_{1},F_{2})/\tau\right] \to \infty$ and $\left[T_{\infty\to1}(F_{1},F_{2})/\sigma\right] \to \infty$ to $\delta_{N} \left[T_{2\to2}(F_{1},F_{2})/\tau\right] \to \infty$ and\\ $\delta_{N} \left[T_{\infty\to1}(F_{1},F_{2})/\sigma\right] \to \infty$ respectively. That is, the infima are taken over a (slightly) larger class of sequences in $H_{1}$. The second difference is the conclusion: the power of the tests no longer tend to one. In fact, the limiting power of the tests may be no greater than $\alpha$. To prove the theorem, the main work is in constructing a sequence of alternatives such that $T_{2\to2}(F_{1},F_{2})/\tau \to 0$ slower than any given sequence, and  $T_{2\to2}(D_{1},D_{2})$ and  $T_{2\to2}(D_{1}^{r},D_{2}^{r})$ converge to the same nondegenerate distribution. Intuitively, Theorem 3 states that the tests proposed in Theorems 1 and 2 cannot detect differences between random graph models that are too similar, in the sense that $T_{2\to 2}(F_{1},F_{2})$ or $T_{\infty\to 1}(F_{1},F_{2})$ are too close to $0$. 


\begin{flushleft}
\textbf{Proof of Theorem 3}: I demonstrate the claim for the test based on the $2\to2$ norm since the proof of that based on the $\infty\to1$ norm is identical. The proof is constructive in that, for any sequence $\delta_{N} \to \infty$, it specifies a specific sequence of distribution function matrices $F_{1}$ and $F_{2}$, depending on $\delta_{N}$, such that $\delta_{N}T_{2\to2}(F_{1},F_{2})/\tau \to \infty$ or
\begin{align*}
\delta_{N}\frac{\max_{s \in \mathbb{R}}\max_{\varphi \in \mathcal{S}^{N}}\sqrt{\sum_{i \in [N]}\left(\sum_{j \in [N]}\left(F_{ij,1}(s) -F_{ij,2}(s)\right)\varphi_{j}\right)^{2}}}{\max_{s \in [S]}\max_{j \in [N]}\sqrt{\sum_{i\in[N]}\left[F_{ij,1}(s) + F_{ij,2}(s) - 2F_{ij,1}(s)F_{ij,2}(s)\right]}} \to \infty
\end{align*}
and 
\begin{align*}
P\left((R+1)^{-1}\left(1 + \sum_{r\in[R]}\mathbbm{1}\{T_{2\to2}(D_{1}^{r},D_{2}^{r}) \geq T_{2\to2}(D_{1},D_{2})\} \right) \leq  \alpha\right) \to \alpha.
\end{align*}
The proof has three steps. The first step is to specify $F_{1}$ and $F_{2}$. For an arbitrary $\varepsilon > 0$, define $A_{1-\varepsilon} = [\lceil\left(1-\varepsilon\right) N\rceil]$ and $A_{\varepsilon} = [N]\setminus A_{1-\varepsilon}$. That is, let $A_{1-\varepsilon}$ index the first $\lceil(1-\varepsilon)N\rceil$ agents in the sample and $A_{\varepsilon}$ the last $\lfloor \varepsilon N \rfloor$. Suppose $F_{ij,1} = F_{ij,2}$ for  $i,j \in A_{1-\varepsilon}$ with $F_{ij,1}$ and $F_{ij,2}$ uniformly bounded away from $0$ and $1$, $F_{ij,1} = F_{ij,2} = 0$ for $i \in A_{\varepsilon}$ and $j \in A_{1-\varepsilon}$ (or $i \in A_{1-\varepsilon}$ and  $j \in A_{\varepsilon}$),  and $F_{ij,1} = 1 + F_{ij,2}$ for $i,j \in A_{\varepsilon}$.\newline\\

The second step is to fix $\varepsilon = \left(\delta_{N}N\right)^{-1/2}$. Since $T_{2\to2}$ is $O\left(N\varepsilon\right)$ and $\tau$ is $O(\sqrt{N})$ by construction from the first step, it follows that $T_{2\to2}(F_{1},F_{2})/\tau \to 0$, but $\delta_{N}T_{2\to2}(F_{1},F_{2})/\tau \to \infty$.  \newline\\

The third step is then to apply the triangle inequality twice. The first application gives
\begin{align*}
&\max_{s \in \mathbb{R}}\max_{\varphi \in \mathcal{S}^{N}}\sqrt{\sum_{i \in [N]}\left(\sum_{j \in [N]}\left(\mathbbm{1}_{D_{ij,1} \leq s} - \mathbbm{1}_{D_{ij,2} 
\leq s}\right)\varphi_{j}\right)^{2}} \\
&\geq \max_{s \in \mathbb{R}}\max_{\varphi \in \mathcal{S}^{N}}\sqrt{\sum_{i \in A_{1-\varepsilon}}\left(\sum_{j \in A_{1-\varepsilon}}\left(\mathbbm{1}_{D_{ij,1} \leq s} - \mathbbm{1}_{D_{ij,2}  \leq s}\right)\varphi_{j}\right)^{2}} 
- \max_{s \in \mathbb{R}}\max_{\varphi \in \mathcal{S}^{N}}\sqrt{\sum_{i \in A_{\varepsilon}}\left(\sum_{j \in A_{\varepsilon}}\left(\mathbbm{1}_{D_{ij,1} \leq s} - \mathbbm{1}_{D_{ij,2}  \leq s}\right)\varphi_{j}\right)^{2}}.
\end{align*}
The second application gives
\begin{align*}
&\max_{r\in[R]}\max_{s \in \mathbb{R}}\max_{\varphi \in \mathcal{S}^{N}}\sqrt{\sum_{i \in [N]}\left(\sum_{j \in [N]}\left(\mathbbm{1}_{D_{ij,1}^{r} \leq s} - \mathbbm{1}_{D_{ij,2}^{r} 
\leq s}\right)\varphi_{j}\right)^{2}} \\
&\leq \max_{r\in[R]}\max_{s \in \mathbb{R}}\max_{\varphi \in \mathcal{S}^{N}}\sqrt{\sum_{i \in A_{1-\varepsilon}}\left(\sum_{j \in A_{1-\varepsilon}}\left(\mathbbm{1}_{D_{ij,1}^{r} \leq s} - \mathbbm{1}_{D_{ij,2}^{r}  \leq s}\right)\varphi_{j}\right)^{2}} + \max_{r\in[R]}\max_{s \in \mathbb{R}}\max_{\varphi \in \mathcal{S}^{N}}\sqrt{\sum_{i \in A_{\varepsilon}}\left(\sum_{j \in A_{\varepsilon}}\left(\mathbbm{1}_{D_{ij,1}^{r} \leq s} - \mathbbm{1}_{D_{ij,2}^{r}  \leq s}\right)\varphi_{j}\right)^{2}}. 
\end{align*}
Both $\max_{s \in \mathbb{R}}\max_{\varphi \in \mathcal{S}^{N}}\sqrt{\sum_{i \in A_{\varepsilon}}\left(\sum_{j \in A_{\varepsilon}}\left(\mathbbm{1}_{D_{ij,1} \leq s} - \mathbbm{1}_{D_{ij,2}  \leq s}\right)\varphi_{j}\right)^{2}} $ and $\max_{r\in[R]}\max_{s \in \mathbb{R}}\max_{\varphi \in \mathcal{S}^{N}}\sqrt{\sum_{i \in A_{\varepsilon}}\left(\sum_{j \in A_{\varepsilon}}\left(\mathbbm{1}_{D_{ij,1}^{r} \leq s} - \mathbbm{1}_{D_{ij,2}^{r}  \leq s}\right)\varphi_{j}\right)^{2}}$ are bounded by $N\varepsilon$ by construction and thus are $o\left(\sqrt{N}\right)$ by the second step. On the other hand, 
$\max_{s \in \mathbb{R}}\max_{\varphi \in \mathcal{S}^{N}}\sqrt{\sum_{i \in A_{1-\varepsilon}}\left(\sum_{j \in A_{1-\varepsilon}}\left(\mathbbm{1}_{D_{ij,1} \leq s} - \mathbbm{1}_{D_{ij,2}  \leq s}\right)\varphi_{j}\right)^{2}}/\sqrt{N}$ and $\max_{r\in[R]}\max_{s \in \mathbb{R}}\max_{\varphi \in \mathcal{S}^{N}}\sqrt{\sum_{i \in A_{1-\varepsilon}}\left(\sum_{j \in A_{1-\varepsilon}}\left(\mathbbm{1}_{D_{ij,1}^{r} \leq s} - \mathbbm{1}_{D_{ij,2}^{r}  \leq s}\right)\varphi_{j}\right)^{2}}/\sqrt{N}$ are bounded away from $0$ by the lower bound in Lemma 1. Since these two are identically distributed and nondegenerate by construction, the result follows. $\square$ \newline \\

\subsection{Pointwise Consistency for Regular Alternatives}

Under certain conditions the power of the tests from Theorems 1 and 2 tend to one whenever the difference between $F_{1}$ and $F_{2}$ is ``regular'' in the sense that there exist two nontrivially sized subcommunities $I_{N}, J_{N} \subseteq [N]$ such that the probabilities that any agent in $I_{N}$ links to any agent in $J_{N}$ all either increase or decrease with $t$. The hypothesis of this theorem does not specify rate conditions that depend on operator norms, and may be easier to interpret and apply in practice. I demonstrate its use with the concrete example of Section 2. 
\vspace{2mm}
\begin{flushleft}
\textbf{Theorem 4}: Suppose there exists $I_{N}, J_{N} \subseteq [N]$ with $\liminf_{N\to\infty} \frac{\left| I_{N}\right| \wedge \left|J_{N}\right|}{N} > 0$, $s \in \mathbb{R}$, and $\rho_{N} > 0$ such that for all $i \in I_{N}$ and  $j \in J_{N}$ either $\left(F_{ij,1}(s) - F_{ij,2}(s) \right)> \rho_{N}$ or $\left(F_{ij,1}(s) - F_{ij,2}(s) \right)< \rho_{N}$. Then the power of the test from Theorem 1 converges to one if $\rho_{N}N/\ln(N) \to \infty$. The power of the test from Theorem 2 converges to one if $\rho_{N}N \to \infty.$ $\square$
\end{flushleft}
\vspace{2mm}
\end{flushleft}

The benefit of Theorem 4 relative to Theorems 1 and 2 is that its hypothesis does not use rate conditions that depend on operator norms. To illustrate its use, I sketch two simple testing problems with models based on the concrete example from Section 2. Recall for this example that $F_{ij,t}(s) = G_{ij,t}(s-f_{t}(\alpha_{i,t},\alpha_{j,t},w_{ij,t}))$. Suppose for example that the idiosyncratic errors are identically distributed for all agent-pairs and networks, the agent-pair attributes are the same for the two networks, and the community link function has the form $f(\alpha_{i,t},\alpha_{j,t},w_{ij}) = \Lambda(\alpha_{i,t} + \alpha_{j,t} + w_{ij}\beta)$ for some unknown vector $\beta$ and strictly monotonic function $\Lambda$. Then the hypothesis of Theorem 4 is satisfied if there exists an $I_{N}$ with $\liminf_{N \to \infty} I_{N}/N > 0$ such that $\left|\alpha_{i,1}-\alpha_{i,2}\right| > \rho_{N}$ for all $i \in I_{N}$. That is, under these conditions, the tests of Theorems 1 and 2 eventually (correctly) reject the null hypothesis that the two networks have the same collection of agent-specific effects when it is false. 

Alternatively, suppose that the idiosyncratic errors are identically distributed, the agent-specific effects and agent-pair attributes are the same for the two networks, and the community link function has the form $f_{t}(\alpha_{i},\alpha_{j},w_{ij}) = \Lambda_{t}(\alpha_{i},\alpha_{j}) + w_{ij}\beta$ for some functions $\{\Lambda_{t}\}_{t\in[2]}$ and vector $\beta$. Then the hypothesis of Theorem 4 is satisfied if $\Lambda_{1}(\alpha_{i},\alpha_{j})$ and $\Lambda_{2}(\alpha_{i},\alpha_{j})$ disagree on $I_{N}\times J_{N}$ with $\liminf_{N \to \infty} (I_{N}\wedge J_{N})/N > 0$. That is, under these conditions, the tests from Theorems 1 and 2 eventually (correctly) reject the null hypothesis that the two networks have the same community link function when it is false.

\begin{flushleft}
\textbf{Proof of Theorem 4}:  The claim is proven by checking the hypotheses of Theorems 1 and 2. This is done in three steps. The first step is to demonstrate that the assumption that there exists $I_{N}, J_{N} \subseteq [N]$ with $\liminf_{N\to\infty} \frac{\left| I_{N}\right| \wedge \left|J_{N}\right|}{N} > 0$ and $\rho_{N} > 0$ such that for all $i \in I_{N}, j \in J_{N}$ there exists $s$ such that $\left(F_{ij,1}(s) - F_{ij,2}(s) \right)> \rho_{N}$ or $\left(F_{ij,1}(s) - F_{ij,2}(s) \right)< \rho_{N}$ implies that $T_{2\to2}(F_{1},F_{2}) \geq \rho_{N}N$ and $T_{\infty\to1}(F_{1},F_{2}) \geq \rho_{N}N^2$ eventually ($N\to \infty$). Write $\delta = \liminf_{N\to\infty} \frac{\left| I_{N}\right| \wedge \left|J_{N}\right|}{N} > 0$. Then 
\begin{align*}
T_{2\to2}(F_{1},F_{2}) &= \max_{s \in \mathbb{R}}\max_{\varphi: ||\varphi||_{2} = 1}\left(\sum_{i\in[N]}\left(\sum_{j \in [N]}(F_{ij,1}(s) - F_{ij,2}(s))\varphi_{j}\right)^{2}\right)^{1/2} \\
&\geq \left(\sum_{i\in I_{N}}\left(\sum_{j \in J_{N}}\frac{(F_{ij,1}(s) - F_{ij,2}(s))}{\sqrt{\sum_{j\in J_{N}}1}}\right)^{2}\right)^{1/2} \geq \rho_{N}N\delta^{2}
\end{align*}
eventually and 
\begin{align*}
T_{\infty\to1}(\rho_{N}F_{1},\rho_{N}F_{2}) &= \max_{s \in \mathbb{R}}\max_{\varphi: ||\varphi||_{\infty} = 1}\sum_{i\in[N]}\left|\sum_{j \in [N]}(F_{ij,1}(s) -F_{ij,2}(s)\varphi_{j}\right| \\
&= \max_{s \in \mathbb{R}}\max_{\varphi: ||\varphi||_{\infty} = 1}\max_{\psi: ||\psi||_{\infty} = 1}\sum_{i\in[N]}\sum_{j \in [N]}(F_{ij,1}(s) - F_{ij,2}(s))\varphi_{j}\psi_{i} \\
&\geq \max_{s \in \mathbb{R}}\max_{\varphi \in \{0,1\}^{N}}\max_{\psi \in \{0,1\}^{N}}\left|\sum_{i\in[N]}\sum_{j \in [N]}(F_{ij,1}(s) - F_{ij,2}(s))\varphi_{j}\psi_{i}\right| \\
&\geq \max_{s\in\mathbb{R}}\left|\sum_{i\in I_{N}}\sum_{j \in J_{N}}(F_{ij,1}(s) - F_{ij,2}(s))\right|
\geq \rho_{N}N^2\delta^{2}
\end{align*}
eventually where the first inequality follows from the fact that for any $N \times N$ dimensional matrix $X$
\begin{align*}
 \max_{\varphi \in \{0,1\}^{N}}\max_{\psi \in \{0,1\}^{N}}&\left|\sum_{i \in [N]}\sum_{j \in [N]}X_{ij}\varphi_{j}\psi_{i}\right| 
 = \max_{\varphi \in \{-1,1\}^{N}}\max_{\psi \in \{-1,1\}^{N}}\left|\sum_{i \in [N]}\sum_{j \in [N]}X_{ij}\left(\frac{\varphi_{j}+1}{2}\right)\left(\frac{\psi_{i}+1}{2}\right)\right|\\
&= \max_{\varphi \in \{-1,1\}^{N}}\max_{\psi \in \{-1,1\}^{N}}\left|\sum_{i \in [N]}\sum_{j \in [N]}\left[X_{ij}\varphi_{j}\psi_{i} + X_{ij} + X_{ij}\varphi_{j} + X_{ij}\psi_{i}\right]/4\right| \\
&\leq \max_{\varphi \in \{-1,1\}^{N}}\max_{\psi \in \{-1,1\}^{N}}\sum_{i \in [N]}\sum_{j \in [N]}X_{ij}\varphi_{j}\psi_{i} = \max_{\varphi: ||\varphi||_{\infty} = 1}\max_{\psi: ||\psi||_{\infty} = 1}\sum_{i \in [N]}\sum_{j \in [N]}X_{ij}\varphi_{j}\psi_{i}. 
\end{align*}
This last inequality for this second block of equations follows from distributing the maximzation over the sum. 
\newline

The second step  is to observe that the assumption that there exists $I_{N}, J_{N} \subseteq [N]$ with $\liminf_{N\to\infty} \frac{\left| I_{N}\right| \wedge \left|J_{N}\right|}{N} > 0$ and $\rho_{N} > 0$ such that for all $i \in I_{N}, j \in J_{N}$ there exists $s$ such that $\left(F_{ij,1}(s) - F_{ij,2}(s) \right)> \rho_{N}$ or $\left(F_{ij,1}(s) - F_{ij,2}(s) \right)< \rho_{N}$ implies that
\begin{align*}
\delta\sqrt{\rho_{N}N} \leq \tau \leq \sqrt{2}\sqrt{\rho_{N}N} \text{ and } \delta\sqrt{\rho_{N}N^3} \leq \sigma \leq \sqrt{2}\sqrt{\rho_{N}N^3}
\end{align*}
eventually. Without loss of generality suppose that $\left(F_{ij,1}(s) - F_{ij,2}(s) \right)> \rho_{N}$. The two upper bounds then follow from the fact that $F_{ij,t}(s)$ is bounded in $[0,1]$. The first lower bound follows from
\begin{align*}
\tau &= \max_{s \in \mathbb{R}}\max_{i \in [N]} \sqrt{\sum_{j \in [N]}\left(F_{ij,1}(s) + F_{ij,2}(s) - 2F_{ij,1}(s)F_{ij,2}(s)\right)} \\
&\geq  \max_{s \in \mathbb{R}}\max_{i \in I_{N}} \sqrt{\sum_{j \in J_{N}}\left(\left[F_{ij,1}(s) - F_{ij,2}(s)\right] + 2F_{ij,2}(s)(1-F_{ij,1}(s))\right)} 
\end{align*}
and the fact that $2F_{ij,2}(s)(1-F_{ij,1}(s))$ is not negative. Similarly
 \begin{align*}
\sigma &= \max_{s \in \mathbb{R}}\sum_{i\in[N]}\sqrt{\sum_{j \in [N]}\left(F_{ij,1}(s) + F_{ij,2}(s) - 2F_{ij,1}(s)F_{ij,2}(s)\right)} \\
&\geq  \max_{s \in \mathbb{R}}\sum_{i \in I_{N}}\sqrt{\sum_{j \in J_{N}}\left(\left[F_{ij,1}(s) - F_{ij,2}(s)\right] + 2F_{ij,2}(s)(1-F_{ij,1}(s))\right)}
\end{align*}
implies the second lower bound. 
\newline

The third step is to observe that steps 1 and 2 imply that 
\begin{align*}
T_{2\to2}(F_{1},F_{2})/\tau \geq \sqrt{\rho_{N}N}\delta^{2}/\sqrt{2} \text{ and } T_{\infty\to1}(F_{1},F_{2})/\sigma \geq \sqrt{\rho_{N}N}\delta^{2}/\sqrt{2}
\end{align*}
eventually so that $T_{2\to2}(F_{1},F_{2})/\tau \to \infty \text{ and } T_{\infty\to1}(F_{1},F_{2})/\sigma \to \infty$ so long as $\rho_{N}N \to \infty$. Since 
\begin{align*}
\sigma/\sqrt{\ln(N)} \geq \delta\sqrt{\rho_{N}N^3/\ln(N)}
\end{align*}
eventually, $\rho_{N}N \to \infty$ also implies that $\sigma/\sqrt{\ln(N)} \to \infty$ and so the hypothesis of Theorem 2 is satisfied. Since
 \begin{align*}
 \tau/\sqrt{\ln(N)} \geq \delta\sqrt{\rho_{N}N/\ln(N)},
 \end{align*}
eventually, strengthening the rate condition to $\rho_{N}N/\ln(N) \to \infty$ implies that $\tau/\sqrt{\ln(N)} \to \infty$ so that the hypothesis of Theorem 1 is also satisfied. This demonstrates the claim. $\square$
\end{flushleft}

\subsection{Bounds on power}
The arguments underlying the proofs of Theorems 1-2 can be modified to provide bounds on the power of the two tests. The upper bounds are relevant to the case in which the effect size is small relative to the reference distribution and the lower bounds are relevant to the case in which the effect size is large relative to the reference distribution. The upper bounds do not depend on $\alpha$ because the (lower) bounds on the reference distribution given by Lemmas 1 and 2 used to construct them hold exactly.


The bounds depend on the parameters $E\left[T_{2\to2}\left(D_{1},D_{2}\right)\right]$, $E\left[T_{\infty\to1}\left(D_{1},D_{2}\right)\right]$, $\tau$, and $\sigma$ as defined in Section 4.2.1 of the main text. The properties of these parameters depend on $F_{1}$ and $F_{2}$. I give the bounds associated with two relatively simple examples in Sections B.3.1 and B.3.2 below.
\vspace{2mm}
\begin{flushleft}\textbf{Theorem 5}: Let $(x)^{2}_{+} = x^{2}\mathbbm{1}_{x > 0}$, $\gamma \in [0,1/2]$ be arbirary, and $K = 1.78$. The power of the $\alpha$-sized test based on $T_{2\to2}$ (given in Theorem 1) is bounded from above by 
\begin{align*}
\exp\left(-\left(\tau -\sqrt[4]{-\frac{N}{2}\ln\left(\frac{\gamma}{N^3}\right)} - E\left[T_{2\to2}\left(D_{1},D_{2}\right)\right]\right)_{+}^{2}/2\right) + \gamma
\end{align*}
and bounded from below by
\begin{align*}
1 - \exp\left(-\left(E\left[T_{2\to2}\left(D_{1},D_{2}\right)\right] -  \sqrt{-2\ln\left(\frac{\alpha}{N}\right)} - (1+ \gamma)2\tau - \frac{6(1+\gamma)}{\sqrt{\ln(1+\gamma)}}\sqrt{\ln (N)}\right)_{+}^{2}/2\right).
\end{align*}
The power of the $\alpha$-sized test based on $S_{\infty\to1}$ (given in Theorem 2) is bounded from above by
\begin{align*}
\exp\left(-\left(\sigma -\sqrt[4]{-\frac{N^{5}}{2}\ln\left(\frac{\gamma}{N^3}\right)} - K E\left[T_{\infty\to1}\left(D_{1},D_{2}\right)\right]   \right)_{+}^{2}/2\right) + \gamma
\hspace{5mm}
\end{align*}
and bounded from below by
\begin{align*}
1 - \exp\left(-\left(E\left[T_{\infty\to1}\left(D_{1},D_{2}\right)\right] - \left( \sqrt{-2\ln\left(\frac{\alpha}{N}\right)} + 2\sigma \right)/K \right)_{+}^{2}/2\right). \hspace{5mm} \square
\end{align*}
\end{flushleft}

\vspace{2mm}
\begin{flushleft}\textbf{Proof of Theorem 5}: The claim follows the logic of Theorems 1 and 2, and so only a sketch is provided here. The probability that either test statistic exceeds the  $1-\alpha$ quantile of its reference distribution is bounded from above following
\begin{align*}
P\left(T\left(D_{1},D_{2}\right) \geq Q_{1-\alpha}\left(D_{1}^{r},D_{2}^{r}\right)\right) &= P\left(U\left(D_{1},D_{2}\right) \geq \left(Q_{1-\alpha}\left(D_{1}^{r},D_{2}^{r}\right) - E\left[T\left(D_{1},D_{2}\right)\right]\right) \right)\\
&\leq \exp\left(-\left(Q_{1-\alpha}\left(D_{1}^{r},D_{2}^{r}\right) - E\left[T\left(D_{1},D_{2}\right)\right] \right)_{+}^{2}/2\right)
\end{align*}
and bounded from below following
\begin{align*}
P\left(T\left(D_{1},D_{2}\right) \geq Q_{1-\alpha}\left(D_{1}^{r},D_{2}^{r}\right)\right) &= 1 - P\left(-U\left(D_{1},D_{2}\right) \geq \left(E\left[T\left(D_{1},D_{2}\right)\right] - Q_{1-\alpha}\left(D_{1}^{r},D_{2}^{r}\right)\right) \right)\\
&\geq 1 - \exp\left(-\left(E\left[T\left(D_{1},D_{2}\right)\right] - Q_{1-\alpha}\left(D_{1}^{r},D_{2}^{r}\right)\right)_{+}^{2}/2\right)
\end{align*}
where $Q_{1-\alpha}\left(D_{1}^{r},D_{2}^{r}\right)$ refers to the $(1-\alpha)$-quantile of $\{T\left(D_{1}^{r},D_{2}^{r}\right)\}_{r \in [R]}$, $U\left(D_{1},D_{2}\right) =  \left(T\left(D_{1},D_{2}\right) - E\left[T\left(D_{1},D_{2}\right)\right] \right)$, $T$ may refer to one of $T_{2\to2}$ or $T_{\infty\to1}$, $(x)^{2}_{+} = x^{2}\mathbbm{1}_{x > 0}$, and the inequality is due to Talagrand \citep*[see][Theorem 6.10]{boucheron2013concentration} since $T_{2\to2}$ and $T_{\infty\to1}$ are both convex Lipschitz functions. 
\newline

Applying the bounds from Lemma 1 to the test based on $T_{2\to2}$ gives 
\begin{align*}
P\left(T_{2\to2}\left(D_{1},D_{2}\right) \geq Q_{2\to2,1-\alpha}\left(D_{1}^{r},D_{2}^{r}\right)\right) 
\leq  \exp\left(-\left(Q_{2\to2,1-\alpha}\left(D_{1}^{r},D_{2}^{r}\right) - E\left[T_{2\to2}\left(D_{1},D_{2}\right)\right] \right)_{+}^{2}/2\right)
\\
\leq  \exp\left(-\left(\hat{\tau} - E\left[T_{2\to2}\left(D_{1},D_{2}\right)\right]\right)_{+}^{2}/2\right)
\end{align*}
and 
\begin{align*}
P\left(T_{2\to2}\left(D_{1},D_{2}\right) \geq Q_{2\to2,1-\alpha}\left(D_{1}^{r},D_{2}^{r}\right)\right) 
\geq 1 - \exp\left(-\left(E\left[T_{2\to2}\left(D_{1},D_{2}\right)\right] -  Q_{2\to2,1-\alpha}\left(D_{1}^{r},D_{2}^{r}\right)\right)_{+}^{2}/2\right) \\
\geq 1 - \exp\left(-\left(E\left[T_{2\to2}\left(D_{1},D_{2}\right)\right] -  \sqrt{-2\ln\left(\frac{\alpha}{N}\right)} - (1+ \gamma)2\tau - \frac{6(1+\gamma)}{\sqrt{\ln(1+\gamma)}}\sqrt{\ln (N)}\right)_{+}^{2}/2\right)
\end{align*}
for any $\gamma \in [0,1/2]$, where $Q_{2\to2,1-\alpha}\left(D_{1}^{r},D_{2}^{r}\right)$ refers to the $(1-\alpha)$-quantile of $\{T_{2\to2}\left(D_{1}^{r},D_{2}^{r}\right)\}_{r \in [R]}$, $\nu_{ij}(s) = F_{ij,1}(s) + F_{ij,2}(s) - 2F_{ij,1}(s)F_{ij,2}(s)$, $\hat{\tau} = \max_{s \in \mathbb{R}}\max_{i \in [N]}\sqrt{\sum_{j \in [N]}\left(\mathbbm{1}_{D_{ij,1} \leq s} - \mathbbm{1}_{D_{ij,2} \leq s}\right)^{2}}$, and $\tau = \max_{s \in \mathbb{R}}\max_{i \in [N]}\sqrt{\sum_{j \in [N]}\nu_{ij}(s)}$.
\newline

Similarly applying Lemma 2 and the inequalities $T_{\infty\to1} \leq S_{\infty\to1} \leq KT_{\infty\to1}$ to the test based on $S_{\infty\to1}$ gives
\begin{align*}
P\left(S_{\infty\to1}\left(D_{1},D_{2}\right) \geq Q_{\infty\to1,1-\alpha}\left(D_{1}^{r},D_{2}^{r}\right)\right)
\leq \exp\left(-\left(Q_{\infty\to1,1-\alpha}\left(D_{1}^{r},D_{2}^{r}\right) - E\left[S_{\infty\to1}\left(D_{1},D_{2}\right)\right] \right)_{+}^{2}/2\right)
\\
\leq \exp\left(-\left(\hat{\sigma} - K E\left[T_{\infty\to1}\left(D_{1},D_{2}\right)\right]   \right)_{+}^{2}/2\right)
\end{align*}
and 
\begin{align*}
P\left(S_{\infty\to1}\left(D_{1},D_{2}\right) \geq Q_{\infty\to1,1-\alpha}\left(D_{1}^{r},D_{2}^{r}\right)\right)
\geq 1 - \exp\left(-\left(E\left[S_{\infty\to1}\left(D_{1},D_{2}\right)\right] -  Q_{\infty\to1,1-\alpha}\left(D_{1}^{r},D_{2}^{r}\right)\right)_{+}^{2}/2\right) \\
\geq 1 - \exp\left(-\left(E\left[S_{\infty\to1}\left(D_{1},D_{2}\right)\right]  -  \sqrt{-2\ln\left(\frac{\alpha}{N}\right)} - 2\sigma \right)_{+}^{2}/2\right)
\end{align*}
where $Q_{\infty\to1,1-\alpha}\left(D_{1}^{r},D_{2}^{r}\right)$ refers to the $(1-\alpha)$-quantile of $\{S_{\infty\to1}\left(D_{1}^{r},D_{2}^{r}\right)\}_{r \in [R]}$, $\nu_{ij}(s) = F_{ij,1}(s) + F_{ij,2}(s) - 2F_{ij,1}(s)F_{ij,2}(s)$, $\hat{\sigma} =  \max_{s \in \mathbb{R}}\sum_{i \in [N]}\sqrt{\sum_{j \in [N]}\left(\mathbbm{1}_{D_{ij,1} \leq s} - \mathbbm{1}_{D_{ij,2} \leq s}\right)^{2}} $, and $\sigma = \max_{s \in \mathbb{R}}\sum_{i \in [N]}\sqrt{\sum_{j \in [N]}\nu_{ij}(s)}$. 
\newline

The last step of the proof is to replace $\hat{\tau}$ with $\tau$ and $\hat{\sigma}$ with $\sigma$ in the upper bounds. To do this, write for any $i \in [N]$, $s \in \mathbb{R}$, and $t > 0$ 
\begin{align*}
P\left(\sqrt{\sum_{j \in [N]}\left(\mathbbm{1}_{D_{ij,1} \leq s} - \mathbbm{1}_{D_{ij,2} \leq s}\right)^{2}} \leq \sqrt{\sum_{j \in [N]}\nu_{ij}(s)} - \sqrt{t}\right)\\
\leq P\left(\sum_{j \in [N]}\left(\mathbbm{1}_{D_{ij,1} \leq s} - \mathbbm{1}_{D_{ij,2} \leq s}\right)^{2} \leq \sum_{j \in [N]}\upsilon_{ij}(s) - t\right)
 \leq \exp\left( \frac{-2t^{2}}{N} \right)
\end{align*}
where the second inequality is due to Hoeffding \citep*[see][Theorem 2.8]{boucheron2013concentration}. The union bound implies
\begin{align*}
P\left(\hat{\tau} \leq \tau - \sqrt{t}\right) \leq N^{3}\exp\left( \frac{-2t^{2}}{N} \right)
\end{align*}
and 
\begin{align*}
P\left(\hat{\sigma} \leq \sigma - N\sqrt{t}\right) \leq N^{3}\exp\left( \frac{-2t^{2}}{N} \right)
\end{align*}
which when combined with the upper bounds from before give
\begin{align*}
P\left(T_{2\to2}\left(D_{1},D_{2}\right) \geq Q_{2\to2,1-\alpha}\left(D_{1}^{r},D_{2}^{r}\right)\right) 
\leq \exp\left(-\left(\hat{\tau} - E\left[T_{2\to2}\left(D_{1},D_{2}\right)\right]\right)_{+}^{2}/2\right)\\
\leq \exp\left(-\left(\tau - \sqrt{t} - E\left[T_{2\to2}\left(D_{1},D_{2}\right)\right]\right)_{+}^{2}/2\right) + N^{3}\exp\left( \frac{-2t^{2}}{N} \right)
\end{align*} 
and 
\begin{align*}
P\left(S_{\infty\to1}\left(D_{1},D_{2}\right) \geq Q_{\infty\to1,1-\alpha}\left(D_{1}^{r},D_{2}^{r}\right)\right)
\leq \exp\left(-\left(\hat{\sigma} - K E\left[T_{\infty\to1}\left(D_{1},D_{2}\right)\right]   \right)_{+}^{2}/2\right)\\
\leq \exp\left(-\left(\sigma - N\sqrt{t} - KE\left[T_{\infty\to1}\left(D_{1},D_{2}\right)\right]\right)_{+}^{2}/2\right) + N^{3}\exp\left( \frac{-2t^{2}}{N} \right)
\end{align*} 
by the law of total probability. The claim follows from choosing $\gamma = N^{3}\exp(-2t^{2}/N)$. $\square$
\end{flushleft}
\vspace{2mm}

What remains is to characterize the parameters $E\left[T_{2\to2}\left(D_{1},D_{2}\right)\right]$, $E\left[T_{\infty\to1}\left(D_{1},D_{2}\right)\right]$, $\tau$, and $\sigma$. Since these are context specific, I consider two examples in the subsections below. The first example is a degree experiment where the treatment increases or decreases the probability that every pair of agents forms a link by a constant amount. The second example is a cluster experiment where the treatment alters agent assignment to cliques or clusters, but does not alter link probabilities directly.

\subsubsection{Example 1: degree experiment}
The first example is a degree experiment. The treatment increases or decreases the probability that every pair of agents forms a link by some fixed amount. In this example, $D_{1}$ and $D_{2}$ are the adjacency matrices corresponding to two unweighted and undirected networks with no loops. That is, $D_{1}$ and $D_{2}$ are symmetric, binary, and hollow matrices. The entries within a network are also identically distributed: $P(D_{ij,t} = 1) = p_{t} \in (0,1)$. Setting $F_{t}(s) = p_{t}$ in the above definitions for this example gives that $\tau = \sqrt{\left(p_{1} + p_{2} - 2p_{1}p_{2}\right)N}$ and $\sigma =  \sqrt{\left(p_{1} + p_{2} - 2p_{1}p_{2}\right)N^{3}}$. Since the matrix norms $T_{2\to2}$ and $T_{\infty\to1}$ are convex, we also have  $E\left[T_{2\to2}\left(D_{1},D_{2}\right)\right] \geq T_{2\to2}\left(E\left[D_{1}\right],E\left[D_{2}\right]\right) = N|p_{1}-p_{2}|$ and $E\left[S_{\infty\to1}\left(D_{1},D_{2}\right)\right] \geq E\left[T_{\infty\to1}\left(D_{1},D_{2}\right)\right] \geq T_{\infty\to1}\left(E\left[D_{1}\right],E\left[D_{2}\right]\right) = N^{2}|p_{1}-p_{2}|$. 

It follows from Theorem 5 that the power of the test based on $T_{2\to2}$ is bounded from above by 
\begin{align*}
\exp\left(-\left(\sqrt{\left(p_{1} + p_{2} - 2p_{1}p_{2}\right)N} - N|p_{1}-p_{2}| + o\left(\sqrt{N}\right)\right)_{+}^{2}/2\right) + o(1) 
\end{align*}
and bounded from below by 
\begin{align*}
1 - \exp\left(-\left(N|p_{1}-p_{2}| - 2\sqrt{\left(p_{1} + p_{2} - 2p_{1}p_{2}\right)N} + o\left(\sqrt{N}\right)\right)_{+}^{2}/2\right). 
\end{align*}
The power of the test based on $S_{\infty\to1}$ is bounded from above by 
\begin{align*}
\exp\left(-\left( \sqrt{\left(p_{1} + p_{2} - 2p_{1}p_{2}\right)N^{3}} - K N^{2}|p_{1}-p_{2}| +o\left(\sqrt{N}\right)   \right)_{+}^{2}/2\right) + o(1)
\end{align*}
and bounded from below by
\begin{align*}
1 - \exp\left(-\left(N^{2}|p_{1}-p_{2}| - 2 \sqrt{\left(p_{1} + p_{2} - 2p_{1}p_{2}\right)N^{3}}/K + o\left(\sqrt{N}\right) \right)_{+}^{2}/2\right).
\end{align*}
Intuitively, the upper (lower) bounds decrease (increase) with the expected entry-wise difference between $D_{1}$ and $D_{2}$ (given by $|p_{1}-p_{2}|$) and increase (decrease) with the variance of the entry-wise difference between $D_{1}$ and $D_{2}$ (given by $\left(p_{1} + p_{2} - 2p_{1}p_{2}\right)$). The remaining terms are asymptotically negligible. 

\subsubsection{Example 2: cluster experiment}
The second example is a cluster experiment where agents belong to one of two groups. Agents within a group are more (or less) likely to form a link than agents across groups. In this experiment, the treatment only changes the agent's group assignments. Unlike the first example, the expected change in the network has mean zero, and so the power of the test comes from the operator norms ``picking out'' the subset of agents that switch from both being in the same group to being in different groups or switch from being in different groups to being in the same group. 

In this example, $D_{1}$ and $D_{2}$ are the adjacency matrices corresponding to two unweighted and undirected networks with no loops. That is, $D_{1}$ and $D_{2}$ are symmetric, binary, and hollow matrices. Agents are assigned to one of two groups in each period as denoted by the variable $Z_{i,t} \in \{1,2\}$. For any time period, all agents in the same group are linked the probability $p$. Agents in different time periods are linked with probability $q$. That is, $P\left(D_{ij,t} = 1 |Z_{i,t}, Z_{j,t}\right) = p\mathbbm{1}_{Z_{i,t} = Z_{j,t}} + q\mathbbm{1}_{Z_{i,t} \neq Z_{j,t}}$. I assume that $ p \neq q$. 

When $t=1$, half of the agents are assigned to group one and the other half are assigned to group two. That is without loss, $Z_{i,1} = 1 + \mathbbm{1}_{i > N/2}$. When $t=2$ agents switch groups independently randomly with probability $\pi \in (0,1)$. Let $\rho = 2\pi(1-\pi)$ denote the probability for any agent-pair $ij$, either $i$ switches groups in time period $2$ or $j$ switches groups, but not both. Then direct calculation of the variances for each of the four possible group combinations yields $\tau = \sqrt{\left(\left(p(1-p) + q(1-q)\right)(1-\rho) + 2\left(p+q-2pq\right)\rho\right)N}$ and $\sigma = \sqrt{\left(\left(p(1-p) + q(1-q)\right)(1-\rho) + 2\left(p+q-2pq\right)\rho\right)N^{3}}$. Intuitively, $\left(p(1-p) + q(1-q)\right)(1-\rho)$ gives the variance of $(D_{ij,1} -D_{ij,2})$ when $|Z_{i,1}-Z_{j,1}| = |Z_{i,2} - Z_{j,2}|$ (neither agents switch or both agents switch) and $2\left(p+q-2pq\right)\rho$ gives the variance of $(D_{ij,1} -D_{ij,2})$ when $|Z_{i,1}-Z_{j,1}| \neq |Z_{i,2} - Z_{j,2}|$ (either one of the agents switch but not both). As in the first example, convexity of the matrix norms gives $E\left[T_{2\to2}\left(D_{1},D_{2}\right)\right] \geq T_{2\to2}\left(E\left[D_{1}\right],E\left[D_{2}\right]\right) = N|p-q|\rho$ and $E\left[T_{\infty\to1}\left(D_{1},D_{2}\right)\right] \geq T_{\infty\to1}\left(E\left[D_{1}\right],E\left[D_{2}\right]\right) = N^{2}|p-q|\rho$. 

It follows from Theorem 5 that the power of the test based on $T_{2\to2}$ is bounded from above by 
\begin{align*}
\exp\left(-\left(2\sqrt{\left(\left(p(1-p) + q(1-q)\right)(1-\rho) + 2\left(p+q-2pq\right)\rho\right)N} - N|p-q|\rho + o\left(\sqrt{N}\right)\right)_{+}^{2}/2\right) + o(1)
\end{align*}
and bounded from below by 
\begin{align*}
 \geq 1 - \exp\left(-\left(N|p-q|\rho - 2\sqrt{\left(\left(p(1-p) + q(1-q)\right)(1-\rho) + 2\left(p+q-2pq\right)\rho\right)N} + o\left(\sqrt{N}\right)\right)_{+}^{2}/2\right).
\end{align*}
The power of the test baed on $S_{\infty\to1}$ is bounded from above by 
\begin{align*}
\exp\left(-\left(2\sqrt{\left(\left(p(1-p) + q(1-q)\right)(1-\rho) + 2\left(p+q-2pq\right)\rho\right)N^{3}} - K N^{2}|p-q|\rho + o\left(\sqrt{N}\right) \right)_{+}^{2}/2\right) + o(1)
\end{align*}
and bounded from below by
\begin{align*}
1 - \exp\left(-\left(N^{2}|p-q|\rho - 2 \sqrt{\left(\left(p(1-p) + q(1-q)\right)(1-\rho) + 2\left(p+q-2pq\right)\rho\right)N^{3}}/K + o\left(\sqrt{N}\right)\right)_{+}^{2}/2\right).
\end{align*}
Intuitively, the bounds increase with the mean of the entry-wise difference between the within and across group linking probabilities (given by $|p-q|$) and the probability that one but not both agents switch groups (given by $\rho = 2\pi(1-\pi)$). It decreases with the variances of the within and across group links (given by $p(1-p)$ and $q(1-q)$), and the variance of their difference (given by $p+q-2pq$). The remaining terms are asymptotically negligible.

\section{Simulation evidence}
Section 4.2 in the main text predicts that the test based on $S_{\infty\to1}$ is potentially more powerful than that based on $T_{2\to2}$ for sparse and degree-heterogeneous alternatives. This subsection provides supporting evidence from two Monte Carlo experiments. It considers the case of unweighted unipartite networks with no loops (symmetric, binary, and hollow adjacency matrices) for simplicity. The purpose of this section is not to simulate data that mimics real-world networks (see instead Section 5 of the main text), but rather to assess the predictions in a controlled environment. Other simulations (not reported) yield qualitatively identical results. 


\subsection{The sparse experiment}
Sparsity is a common feature of social and economic networks. For example, in many social surveys it is common for agents to report less than a dozen connections. To examine the impact of network sparsity on the power of the two tests, I consider two Erd\"os-Renyi graph models. In these models, the adjacency matrices are $\{0,1\}$-valued with $P\left(D_{ij,t} = 1\right) = 1-F_{ij,1}(0) = \frac{8}{N}$ and $1-F_{ij,2}(0) = \frac{5}{N}$ for every $i,j \in [N]$. Agents in the first network have approximately 60\% more links than agents in the second network, violating $H_{0}$. Applying the two tests to data simulated from the models with $N = 50/100$ and $R = 10,000$ yields an average p-value for the test based on the $2\to2$ norm of approximately $0.070/0.020$ and an average p-value for the test based on the $\infty\to1$ norm of approximately $0.049/0.013$. The test based on the $\infty\to1$ norm is more powerful, but not dramatically so. 

\subsection{The degree heterogeneous experiment}
Degree heterogeneity is another common feature of social and economic networks. For example, in many production and collaboration networks it is common for a small number of agents to have an order of magnitude more links than the median agent. To examine the impact of degree heterogeneity on the power of the two tests, I consider two second-order stochastic blockmodels. In these models, $P(D_{ij,t} = 1) = 1-F_{ij,t}(0)$, with $F_{1j,1}(0) = F_{1j,2}(0) = .5$ for all $j \in [N]$ and  $1-F_{ij,1}(0) = .02$ and $1-F_{ij,2}(0) = .08$ for any $i,j \in [N]\setminus[1]$. Agents in the first network have approximately $400\%$ percent more links than in the second network, violating $H_{0}$. However, the high degree agent, agent 1, has approximately the same number of links. Applying the two tests to data simulated from the models with $N = 50/100$ and $R = 10,000$ yields an average p-value for the test based on the $2\to2$ norm of approximately $0.521/0.204$ and an average p-value for the test based on the $\infty\to1$ norm of approximately $0.001/0.000$. The test based on the $\infty\to1$ norm is substantially more powerful.

\section{Details about applications and extensions}
\subsection{Application 1: a test of link stationarity}
 \cite{goyal2006economics} observe co-authorships between economists over time and argue that the profession has become more interconnected in response to new research technologies such as the internet. The framework of Section 2 can be used to evaluate whether the changes in network structure are statistically significant. Let $D_{ij,t}$ describe the existence of a co-authorship between economists $i$ and $j$ in time period $t$. Suppose that the researcher observes the co-authorship data for $M$ time periods. Then $H_{0}: F_{1} = F_{2} = ... = F_{M}$ is the hypothesis of link stationarity that the differences between co-authorships over time is explained by $M$ draws from the same link formation model.
 
 One can extend the randomization test of Section 3 to this testing problem by independently permuting all of the $M$ links associated with each agent-pair. For the choice of test statistic, I recommend the maximum or average difference over all $M \choose 2$ pairs of networks using the semidefinite approximation to the $\infty\to1$-norm: $\max_{t, t' \in [M]}S_{\infty\to1}\left(D_{t},D_{t'}\right)$ or $\sum_{t, t' \in [M]}S_{\infty\to1}\left(D_{t},D_{t'}\right)$. The test proposed in the hypothesis of Theorem 2 corresponds to the case of $M = 2$. Theorem 2 applies to the case of $M> 2$ mutatis mutandis.
 
 Failure to reject $H_{0}$ using the network data $D_{1}, D_{2}, ... , D_{M}$ suggests that the observed changes in network interconnectedness are not statistically significant. The first example in Section 5 demonstrates this application to testing link stationarity. 
 
 \subsection{Application 2: a test for link heterogeneity}
\cite{banerjee2013diffusion} collect data on a dozen social and economic ties between villagers in Karnataka, India. \cite{jackson2017economic} suggest that this data on multiple types of connections between villagers ``encode richer information than simply identifying whether two people are close or not.''  The framework of Section 2 can be used to evaluate this hypothesis. Let $D_{ij,1}$ denote whether agents $i$ and $j$ have one type of connection (they report being friends) and $D_{ij,2}$ denote whether agents $i$ and $j$  have another type of connection (they report having borrowed or lent money to one another). Then $H_{0}: F_{1} = F_{2}$ is the hypothesis of link homogeneity that the differences between the networks can be explained by draws from the same link formation model. Failure to reject $H_{0}$ using the network data $D_{1}$ and $D_{2}$ suggests that the observed differences between the friendship and lending networks are not statistically significant. The second example in Section 5 demonstrates this application to testing link homogeneity. 

\subsection{Application 3: a test of no treatment effects}
\cite{rose2004we} analyzes yearly aggregate international trade data and argues that participation in trade agreements such as the World Trade Organization (WTO) does not significantly alter the level of trade between countries. The framework of Section 2 can be used to evaluate this hypothesis. Let $D_{ij,t}$ describe the logarithm of the total value of trade between countries $i$ and $j$ in year $t$, and $X_{ij,t}$ be an indicator for whether country $i$ or country $j$ are members of the WTO in year $t$. Let $N$ denote the number of countries and $M$ denote the number of time periods. A nonparametric version of \cite{rose2004we}'s gravity model of trade is
\begin{align*}
D_{ij,t} = \alpha_{i,t} + \alpha_{j,t} + \beta_{ij} + \gamma_{ij,t}X_{ij,t} + \varepsilon_{ij,t}
\end{align*}
 where $\alpha_{i,t}$ and $\alpha_{j,t}$ are country-specific determinants of trade that may vary over time such as GDP or population, $\beta_{ij}$ are country-pair-specific determinants of trade that do not vary over time such as physical distance, and $\varepsilon_{ij,t}$ is an independent and identically distributed idiosyncratic error. The null hypothesis of no treatment effects is $H_{0}: \gamma_{ij,t} = 0$ for all $i,j \in [N]$ and $t \in [M]$. \cite{rose2004we} also allows for two observed country-pair determinants of trade that vary over time: indicators for whether the two countries share the same currency or one is a colony of the other. One can restrict the randomization to be conditional on the value of these binary variables.

The hypothesis $H_{0}$ can be tested using the framework of Section 2 by taking triple differences so that 
\begin{align*}
\left(\left[(D_{ij,t}- D_{i1,t}) - (D_{2j,t} - D_{21,t})\right] - \left[(D_{ij,t'}- D_{i1,t'}) - (D_{2j,t'} - D_{21,t'})\right]\right)\hspace{30mm} \\
= \left(\left[(\gamma_{ij,t}X_{ij,t} - \gamma_{i1,t}X_{i1,t}) - (\gamma_{2j,t}X_{2j,t} - \gamma_{21,t}X_{21,t})\right] - \left[(\gamma_{ij,t'}X_{ij,t'}- \gamma_{i1,t'}X_{i1,t'}) - (\gamma_{2j,t'}X_{2j,t'} - \gamma_{21,t'}X_{21,t'})\right]\right) \\
+ \left(\left[(\varepsilon_{ij,t} - \varepsilon_{i1,t}) - (\varepsilon_{2j,t} - \varepsilon_{21,t})\right]  - \left[(\varepsilon_{ij,t'}- \varepsilon_{i1,t'}) - (\varepsilon_{2j,t'} - \varepsilon_{21,t'})\right]\right)
\end{align*}
for every $t,t' \in [M]$ and $i,j,\in [N]$. That is, $H_{0}$ implies that $G_{ij,t} = G_{ij,t'}$ for every $t,t' \in [M]$ and $i,j,\in [N]$ where $G_{ij,t}$ refers to the marginal distribution of $\left[(D_{ij,t}- D_{i1,t}) - (D_{2j,t} - D_{21,t})\right]$. This implication can be tested using the framework of Sections 3 and 4, applied to the differenced data  $\left[(D_{ij,t}- D_{i1,t}) - (D_{2j,t} - D_{21,t})\right]$  instead of $D_{ij,t}$. Failure to reject $H_{0}$ using this data suggests that the observed differences in trade over time can be explained by the gravity model of trade with no treatment effects in that any differences in trade for country-pairs across years with and without participation in the WTO are not statistically significant. 


\subsection{Application 4: a test for endogenous link formation}
 \cite{goldsmith2013social} consider a joint model of student GPA and link formation in a high-school social network.\footnote{See also \cite{hsieh2014social,johnsson2015estimation,arduini2015parametric,auerbach2019identification}} They hypothesize that a determinant of GPA also drives variation in network links, and propose a one-sample parametric test for such endogenous link formation. The framework of Section 2 can be used to specify a nonparametric two-sample test. 
 
Let $D_{ij,t}$ denote whether students $i$ and $j$ report a friendship in school-year $t$ and $\eta_{i,t}$ describe the social characteristic of agent $i$ in time period $t$ thought to drive link formation. For example, $\eta_{i,t}$ might be a measure of agent $i$'s participation in an extracurricular activity or membership in a social clique. In the setting of \cite{goldsmith2013social}, $\eta_{i,t}$ is the residual from a linear-in-means model of network peer effects. Network endogeneity may then refer to the idea that the distribution of $D_{ij,t}$ varies with the social proximity of the agents in the social characteristics space as measured by $|\eta_{i,t}-\eta_{j,t}|$. Define 
\begin{align*}
D_{ij,1}^{\dagger} = D_{ij,1}\mathbbm{1}_{|\eta_{i,1}-\eta_{j,1}| > |\eta_{i,2}-\eta_{j,2}|} + D_{ij,2}\mathbbm{1}_{|\eta_{i,1}-\eta_{j,1}| < |\eta_{i,2}-\eta_{j,2}|}\hspace{5mm} \text{ and } \\
D_{ij,2}^{\dagger} = D_{ij,2}\mathbbm{1}_{|\eta_{i,1}-\eta_{j,1}| > |\eta_{i,2}-\eta_{j,2}|} + D_{ij,1}\mathbbm{1}_{|\eta_{i,1}-\eta_{j,1}| < |\eta_{i,2}-\eta_{j,2}|}.
\end{align*}
In words, $D_{ij,1}^{\dagger}$ is an indicator for whether agents $i$ and $j$ are linked when they are (relatively) farther apart in the social characteristics space and $D_{ij,2}^{\dagger}$ is an indicator for whether agents $i$ and $j$ are linked when they are (relatively) closer in the social characteristics space. Let $F_{ij,t}$ refer to the marginal distribution of $D_{ij,t}^{\dagger}$, conditional on the collection of social characteristics $\eta_{1}$ and $\eta_{2}$.

Then $H_{0}: F_{1} = F_{2}$ is the hypothesis of exogenous link formation that the differences in friendship links across school-years are unrelated to student proximity in the social characteristic space. Failure to reject $H_{0}$ using the network and social characteristic data suggests that any relationship between the students' social proximity and the formation of network links is not statistically significant. 

\subsection{Application 5: a test of link reciprocity}
\cite{calvo2009peer} specify a model of network peer effects in which any nomination of a friendship from one agent to another indicates a social tie between agents. The framework of Section 2 can be used to detect potential asymmetries in link nominations. Let  $D_{ij,1} = D_{ji,1}$ be an indicator for whether $i$ nominates $j$ and $D_{ij,2} = D_{ji,2}$ be an indicator for whether $j$ nominates $i$ when surveyed. 

To illustrate the application, suppose that, in contrast to the null hypothesis of nomination symmetry, high out-degree agents (agents who make many nominations) are thought to nominate differently than low out-degree agents (agents who make few nominations). This choice of network statistic is arbitrary: any other network statistic can be used as a substitute for out-degrees to construct the test. Let $N_{i} = \sum_{j}D_{ij,1}$ describe the out-degree of agent $i$. Define
\begin{align*}
D_{ij,1}^{\dagger} = D_{ij,1}\mathbbm{1}_{N_{i} > N_{j}} + D_{ij,2}\mathbbm{1}_{N_{i} < N_{j}} \text{ and } \\
D_{ij,2}^{\dagger} = D_{ij,1}\mathbbm{1}_{N_{i} < N_{j}} + D_{ij,2}\mathbbm{1}_{N_{i} > N_{j}}.
\end{align*}
In words, $D_{ij,1}^{\dagger}$ is an indicator for whether $i$ nominates $j$ when $i$ makes more nominations than $j$ or $j$ nominates $i$ when $j$ makes more nominations than $i$. $D_{ij,2}^{\dagger}$ is an indicator for whether $i$ nominates $j$ when $i$ makes less nominations than $j$ or $j$ nominates $i$ when $j$ makes less nominations than $i$. Let $F_{ij,t}$ refer to the marginal distribution of $D_{ij,t}^{\dagger}$. In contrast to the testing problem in Application 4, the distribution of $F_{ij,t}$ is not conditional on $N_{i}$ and $N_{j}$. That is, when constructing the randomization test for this application, the number of nominations are to be recomputed with each simulation. 

Then $H_{0}: F_{1} = F_{2}$ is the hypothesis of link reciprocity that the differences in nominations between pairs of agents are explained by draws from the same link formation model. Failure to reject $H_{0}$ using the data $D_{1}$ and $D_{2}$ suggests that any asymmetry in linking behavior (for example, associated with the agent out-degrees) is not statistically significant.

\subsection{Application 6: a test for network externalities}
\cite{pelican2020optimal} consider a model of link formation in which the propensity for agents to form a link may depend on the existence of other links  in the network. They propose a one-sample parametric test for such network externalities. The framework of Section 2 can be used to specify a nonparametric two-sample test. 

I illustrate the application with the following nonparametric version of a model motivated by \cite*{bloch2007formation} \citep*[see][Section 2]{graham2015methods}
\begin{align*}
D_{ij,t} = \mathbbm{1}\left\{\alpha_{ij} + \gamma_{ij}\sum_{k=1}^{N}D_{ik,t}D_{jk,t} - \varepsilon_{ij,t} \geq 0\right\}
\end{align*}
where $\varepsilon_{ij,t}$ is independent, identically distributed, and mean-zero. The parameters $\alpha_{ij}$ and $\gamma_{ij}$ do not vary with $t$. In this model, agents with many friends in common are more likely to become friends, and the agents  first draw $\{\varepsilon_{ij,t}\}_{i\neq j}$ and then choose links so that the link formation rule is satisfied for every $ij$-pair. The use of $\sum_{k=1}^{N}D_{ik,t}D_{jk,t}$ on the right-hand side is arbitrary and can be replaced by any network statistic. 

The hypothesis of no network externalities corresponds to $H_{0}: \gamma_{ij} = 0$ for all $i,j \in [N]$. Under this hypothesis, $D_{1}$ and $D_{2}$ are drawn from the same random graph model (in the sense of Section 2.1), and so the randomization test proposed in Section 3 controls size in finite samples.

When $\alpha_{ij}$ and $\gamma_{ij}$ are thought to also vary across the two networks, the researcher might instead test the more general  hypothesis $H_{0}: \alpha_{ij,1} = \alpha_{ij,2}$ and  $\gamma_{ij,1} = \gamma_{ij,2}  = 0 $ for every  $i,j \in  [N]$. Failure to reject $H_{0}$ using the network data $D_{1}$ and $D_{2}$ suggests that the distribution of network links can be explained by draws from a model without network externalities and so any network externalities are not statistically significant.


\subsection{Extension 1: a completely randomized experiment}
\cite{banerjee2018changes} collect data on social connections between villagers in 75 villages before and after a microfinance agency offers loans to villagers in 43 of the villages. They find that villages in which the microfinance agency entered were associated with relatively lower densities and argue that access to microfinance disincentivized the formation of some types of connections in the network. The framework of Section 2 can be extended to test the hypothesis that the changes in the network structure for the treatment villages are statistically significant.   

Let $D_{ij,t,v}$ describe whether villagers $i$ and $j$ in village $v$ report a social connection in time period $t$, $X_{v}$ be a binary indicator for whether village $v$ was assigned to the treatment group, $N_{v}$ represent the number of agents in village $v$, $V_{1}$ be the number of treatment villages, $V_{0}$ be the number of control villages, and  $V = V_{1} + V_{0}$ be the total number of villages. Following \cite{banerjee2018changes}, each villager is assigned to exactly one village, each village is assigned to one of two treatment statuses, villagers do not form social connections across villages, and there are two time periods. Time period $t=1$ is the before period in which no treatment has been assigned to either the treatment or control villages. Time period $t=2$ is the after period in which treatment has been assigned to the treatment ($X_{v} = 1$) but not the control ($X_{v} = 0$) villages. 

To illustrate the extension, I suppose that the microfinance agency selected $V_{1}$ villages uniformly at random from the collection of $V$ villages for treatment. This treatment assignment mechanism corresponds to a ``completely randomized experiment'' in the terminology of \cite{imbensRubin2015}. Different treatment assignment mechanisms imply different strategies for inference. The hypothesis to be tested is the null of $H_{0}: D_{ij,t,v}(0) = D_{ij,t,v}(1)$ for every $i,j \in [N_{v}]$, $v \in [V]$, and $t \in [2]$ \citep[see][Chapter 5]{imbensRubin2015}, where $D_{ij,t,v}(\tau)$ is the potential outcome (network) associated with treatment ($\tau = 1$) or no treatment ($\tau = 0$). 

For this problem, a test statistic is any real-valued function of the collection village adjacency matrices 
\begin{align*}
T\left(\{D_{t,v}\}_{t\in [T], v \in [V]: X_{v} = 1}, \{D_{t,v}\}_{t\in [T], v \in [V]: X_{v} = 0}\right).
\end{align*}
Since the villages may all be defined on communities of different sizes, it is assumed that the test statistic is well-defined on matrices of arbitrary dimension. Test statistics based on the usual network statistics such as density, clustering, eigenvector centrality, etc. satisfy this property. For the reasons outlined in the main text of the paper, I recommend choosing $T$ to be the average squared difference between the entry-wise differences of the two adjacency matrices as measured by the semidefinite approximation to the $\infty\to1$ norm between the treatment and control groups. That is, 
\begin{align*}
T\left(\{D_{t,v}\}_{t\in [T], v \in \{[V]: X_{v} = 1\}}, \{D_{t,v}\}_{t\in [T], v \in \{[V]: X_{v} = 0\}}\right) \\
= \sum_{v \in \{[V]: X_{v} = 1\}, v' \in \{[V]: X_{v'} = 0\}} \left(S_{\infty\to1}(D_{1,v},D_{2,v}) -  S_{\infty\to1}(D_{1,v'},D_{2,v'}) \right)^{2}.
\end{align*}
An appealing feature of this test statistic is that it has a difference-in-differences-like structure in that it measures the difference in the change in the network for the treated villages relative to the change in the control villages. The use of average squared loss here to compare treatment and control villages is arbitrary: what is  important is the use of the $S_{\infty\to1}$ to measure changes within a village over time. 

To construct a critical value for this test, the standard is to generate a reference distribution by re-randomizing the treatment assignment $X_{v}$ \citep[see][Chapter 5]{imbensRubin2015}. That is, let $\mathcal{X} := \{x \in \{0,1\}^{V}: ||x||_{1} = V_{1}\}$ describe the set of possible counterfactual treatment assignments (the set of all subsets of $[V]$ of size $V_{1}$) and let $\{X^{r}\}_{r \in [R]}$ be $R$ independent and uniformly distributed draws from $\mathcal{X}$. Then the standard is to use  
\begin{align*}
\{T\left(\{D_{t,v}\}_{t\in [T], v \in [V]: X^{r}_{v} = 1}, \{D_{t,v}\}_{t\in [T], v \in [V]: X^{r}_{v} = 0} \right)\}_{r \in [R]}
\end{align*}
as a reference distribution for testing $H_{0}$. The construction of the test based on this reference distribution follows exactly Section 3 in the main text. The test controls size by construction. I suspect that it is straightforward to derive power properties of the test based on the $\infty\to1$ norm suggested above using the arguments of Theorems 2, 3 and 5, but leave this to future work.

\subsection{Extension 2: a one-sample test of independence}
\cite{fafchamps2007risk} study link formation in a risk-sharing network and argue that the surveyed network connections are unrelated to the respondents' occupations. The framework of Section 2 can be extended to test that the network links and the agent occupations are unrelated in the following sense. Let $D_{ij}$ describe whether agents $i$ and $j$ report a network connection, $X_{i}$ describe the occupation of agent $i$, and $N$ be the number of agents in the community. Then the hypothesis to be tested is that $\{D_{ij}\}_{i,j \in [N]}$ and $\{(X_{i},X_{j})\}_{i,j \in [N]}$ have mutually independent entries. 

Let $T(D,X)$ be a real-valued test statistic defined on the matrix of network connections and vector of occupation assignments. For this testing problem, I suggest a randomization test based on re-randomizing the occupation assignments. Let $\Pi$ be the set of permutations on $[N]$ and $\{\pi^{r}\}_{r \in [R]}$ be a collection of independent and uniformly distributed draws from $\Pi$.  In words, $\pi_{i}^{r}$ refers to another agent in the community that is randomly assigned to agent $i$ and $X_{i}^{r} := X_{\pi^{r}_{i}}$ refers to the occupation associated with agent $\pi_{i}^{r}$. Then under the null hypothesis, $T(D,X)$ and $T(D,X^{r})$ have the same distribution and so $\{T(D,X^{r})\}_{r \in [R]}$ can be used as a reference distribution to test the null hypothesis, exactly as in Section 3 of the main text.

For the reasons outlined in the main text of the paper, I recommend choosing $T$ to be the semidefinite approximation to the $\infty\to1$ norm of the Hadamard (entrywise) product of the matrices $D$ and $W := \{\mathbbm{1}_{X_{i} = X_{j}}\}_{i,j \in [N]}$. That is,
\begin{align*}
S_{\infty}(D\cdot W,0) =  \frac{1}{2}\max_{s \in \mathbb{R}}\max_{X \in \mathcal{X}_{2N}}\left<\begin{bmatrix} 0_{N\times N} & D\cdot W \\ D\cdot W & 0_{N\times N}  \end{bmatrix},X\right>
\end{align*}
where $\cdot$ refers to the Hadamard product and $0_{N\times N}$ is an $N \times N$ matrix of $0$s. The intuition behind this test statistic is that if agents with similar occupations are more likely to form a link, then one would expect $D\cdot W$ to be larger (in the sense that its matrix norm is bigger) than $D\cdot W^r$, where $W^r :=  \{\mathbbm{1}_{X_{i}^{r} = X_{j}^{r}}\}_{i,j \in [N]}$ is the matrix of occupations assigned at random.  The test controls size by construction. A study of the power properties is left to future work. 

\subsection{Extension 3: a one-sample specification test}
\cite{jackson2007meeting} argue that real-world social networks are connected in ways that are poorly approximated by, for example, an Erd\"os-Renyi model of link formation. The framework of Section 2 can be extended to test whether some observed network data can be explained by a particular parametric model of link formation. Let $D_{ij}$ be the observed networks links described by the (true) unknown model $F_{ij,1}$. Let $F_{ij,2}$ be the distribution of links associated with a model chosen by the researcher. Then $H_{0} : F_{ij,1} = F_{ij,2}$ is the hypothesis that the distribution of links given by the model that generated the data and those given by the model chosen by the researcher are the same. The network formation model $F_{2}$ does not need to have a closed form representation (it only needs to be simulatable).

To extend the framework of Section 2 to this testing problem, I propose converting it to two-sample problem by first drawing network data $D'$ from $F_{2}$. Let $T(D,D')$ be an arbitrary real-valued test statistic as described in Section 3 evaluated on $D$ and $D'$. Then I propose constructing a reference distribution for $T(D,D')$ by drawing additional simulations from $F_{2}$. That is, I suggest independently drawing $2R$ collections of networks from $F_{2}$, collecting them into two groups $\{D^{r}\}_{r \in [R]}$ and $\{D'^{r}\}_{r \in [R]}$, and using $\{T(D^{r},D'^{r})\}_{r \in [R]}$ as a reference distribution for $T(D,D')$. As motivated in the main text, I propose using $S_{\infty\to1}(D,D')$ for the choice of test statistic. The test controls size by construction. I leave a study of the power properties to future work. 

In many cases, the network formation model $F_{2}$ chosen by the researcher may depend on unknown parameters. For example, the researcher may hypothesize that $D$ is drawn from an Erd\"os-Renyi model with some unknown value of link probability $\theta \in [0,1]$. One way to test this hypothesis is to test each parameter individually (or each parameter in a representative subcollection) and reject the null hypothesis only if the test rejects at every value. This test controls size by construction, but may have low power. Another way to test this hypothesis is to estimate the parameters of the model under $H_{0}$ by some method, and use the estimated value of the parameters to specify $F_{2}$ in the above test. I suspect that under certain conditions such a test will control size asymptotically, but leave this to future work.

\bibliographystyle{chicago}
\bibliography{literature}